\documentclass[opre,nonblindrev]{informs3_homepage} 

\usepackage{amssymb,amsfonts,amsmath,algorithm,algpseudocode,url,color,multirow,booktabs,ctable,float,longtable}
\usepackage{setspace}

\definecolor{pastelBlue}{rgb}{0.0,0.4,0.7}
\usepackage{hyperref}
\hypersetup{
colorlinks=true,
allcolors = pastelBlue  
}

\OneAndAHalfSpacedXI 

\usepackage{natbib}
 \bibpunct[, ]{(}{)}{,}{a}{}{,}%
 \def\bibfont{\small}%
 \def\bibsep{\smallskipamount}%
 \def\bibhang{24pt}%
\TheoremsNumberedThrough     
\ECRepeatTheorems

\EquationsNumberedThrough    

\newcommand{\no}{\nonumber}
\newcommand{\mb}{\mathbb}
\newcommand{\mrm}{\mathrm}
\def\dv{d_{TV}}
\def\C{\mathcal{C}}
\def\le{\left}
\def\ri{\right}

\def\P{\mathbb P}
\def\E{\mathbb E}
\def\1{\mathbb 1}

\def\ind{{\mathbb I}}
\def\gens{{\mathbb G}}
\def\subG{{\mathbb H}}

\def\deg{\textrm{deg}}
\def\alg{{\sf RG}}
\def\algname{{\sf RandGraph}}
\def\bipalgname{{\sf BipRandGraph}}
\def\unif{{\sf U}}
\def\true{\textrm{true}}
\def\sB{\mathbb{B}}
\def\sD{\mathbb{D}}

\def\nedge{m}
\def\bQ{\mathbf{Q}}
\def\bM{\mathbf{M}}
\def\bP{\mathbf{P}}
\def\tf{\sf{TF}}

\begin{document}



\RUNTITLE{Random networks}

\TITLE{Generating Random Networks Without Short Cycles}

\ARTICLEAUTHORS{%
\AUTHOR{Mohsen Bayati}
\AFF{Graduate School of Business, Stanford University, Stanford, CA 94305, \EMAIL{bayati@stanford.edu}}
\AUTHOR{Andrea Montanari}
\AFF{Departments of Electrical Engineering and Statistics, Stanford, CA 94305, \EMAIL{montanar@stanford.edu}}
\AUTHOR{Amin Saberi}
\AFF{Departments of Management Science and Engineering and
Institute for Computational and Mathematical Engineering, Stanford, CA 94305, \EMAIL{saberi@stanford.edu}}
} 

\ABSTRACT{
Random graph generation is an important tool for studying large complex networks. Despite abundance of random graph models, constructing models with application-driven constraints is poorly understood. In order to advance state-of-the-art in this area, we focus on random graphs without short cycles as a stylized family of graphs, and propose the $\algname$ algorithm for randomly generating them. For any constant $k$, when $m=O(n^{1+1/[2k(k+3)]}\,)$, $\algname$ generates an asymptotically uniform random graph with $n$ vertices,  $m$ edges, and no cycle of length at most $k$ using $O(n^2m)$ operations. We also characterize the approximation error for finite values of $n$. To the best of our knowledge, this is the first polynomial-time algorithm for the problem. $\algname$ works by sequentially adding $m$ edges to an empty graph with $n$ vertices. Recently, such sequential algorithms have been successful for random sampling problems. Our main contributions to this line of research includes introducing a new approach for sequentially approximating edge-specific probabilities at each step of the algorithm, and providing a new method for analyzing such algorithms.
}

\KEYWORDS{Network models, Poisson approximation, Random graphs}

\maketitle

%


%
%
\section{Introduction}\label{sec:intro}

Recently, a common objective in many application areas has been extracting information from data sets that contain a network structure. Examples of such data are the Internet, social networks, biological networks, or healthcare networks such as network of physician referrals. In the last example, consider the question ``how is the network of physician referrals formed?''. Answering this question could allow policy makers to influence the formation of the network with the objective of improving quality of care. This could be achieved by rewarding referrals to higher quality physicians and penalizing referrals to lower performing physicians. Unfortunately, empirical analysis of such network related questions is challenging since in most cases researchers have access to a single network or a few snapshots of it over time. Specifically, the small number of samples renders the estimation part of any parametric network formation model unreliable \citep{Chandrasekhar}.

A popular approach in statistical data analysis, when facing small number of observations, is bootstrap \citep{efron79} which increases the number of observations by creating random re-samples of the original data. However, creating random copies of networks can be computationally expensive. For example, if the aim is to create a random copy of the physician referral network while keeping the number of neighbors (degree) of each node fixed, the problem becomes NP hard in general \citep{Wormald1999}. The property of fixing the number of neighbors is relevant when it is desired to control for variations in abilities of the physicians to form working relationships. Similarly, one could be interested in creating random copies of a network when certain sub-structures should be preserved or avoided.
This problem in general is unsolved from a theoretical point of view except for few examples where efficient algorithms are proposed \citep{Wormald1999}. Therefore, practitioners use non-rigorous heuristic models of random networks which may lead to incorrect (biased) estimates, see \citep{MiloShenOrrItzkovitzKashtanChklovskiiAlon} for such a heuristic.

The objective of this paper is to advance state-of-the-art
in this line of research by proposing a new algorithm and analysis technique. We present the approach for a stylized
subclass of problems, generating random graphs without short cycles, and leave extensions to other substructures for
future research. While our emphasis in this paper is on advancing the methodology, and the family of graphs without short cycles is selected
as an example of open problems in this area, we note that randomly generating graphs from this family has practical implications in information theory. Such graphs are used in designing low density parity check (LDPC) codes that can achieve Shannon capacity for transmitting messages in a noisy environment \citep{MCT}.

\subsection{Contributions}

We present a simple and efficient algorithm, $\algname$, for randomly generating simple graphs
without short cycles. 
For any constant $k$,  $\alpha\leq1/[2k(k+3)]$, and $m=O(n^{1+\alpha})$, $\algname$ generates an asymptotically uniform random graph with $n$ vertices,  $m$ edges, and no cycle of length $k$ or smaller. $\algname$ uses $O(n^2m)$ operations in expectation. In addition, for finite values of $n$, we calculate the approximation error.  To the best of our knowledge, this is the first polynomial-time algorithm for the problem.

$\algname$ starts with an empty graph and sequentially adds $m$ edges between pairs of non-adjacent
vertices. In every step, two distinct vertices $i$, $j$ with distance at least $k$ are selected with probability $p_{ij}$, and the edge $(ij)$ is added to the graph. The most crucial step, computing $p_{ij}$, is obtained by finding a sharp estimate for the number of extensions of the partially constructed graph, $G_t$, that contain $(ij)$ and have no cycle of length at most $k$. This estimation is done by computing the expected number of small cycles produced if the rest of the edges are added uniformly at random, using a Poisson approximation.

Our analysis of $\algname$ involes three approximation steps. First we approximate random graphs that have $m$ edges and $n$ vertices with Erd\"{o}s-R\'{e}nyi (ER) graphs where each edge appears independently with probability $m/{n\choose 2}$. The second approximation uses Janson inequality \citep{Janson90} for estimating the probability\footnote{We note that using the Poisson approximation method in \S 6.2 of \citep{JLR00} one can estimate this probability with an additive error that converges to $0$ with a rate that is inversely polynomial in $n$. However, here we require a stronger approximation since we need a multiplicative error that converges to $1$. This would require the additive error to converge to zero faster than the probability of the event itself which is exponentially small in $n$ when $m=O(n^{1+\alpha})$.} that random ER graphs have no cycle of length at most $k$. These two approximations provide us with an estimate for the uniform distribution on the family of graphs without cycles of length at most $k$. In the final and third step, we approximate $G_t$ with ER graphs with edge density $t/m$ to estimate the output distribution of $\algname$, and to show that it is asymptotically equal to the uniform distribution.
We emphasize that these approximations are easy when $m=O(n)$, and our main contribution is to show that they are sharp even
when the number of edges is super-linear in $n$, namely when $m=O(n^{1+\alpha})$ for small values of $\alpha$.

We also provide a theoretical and empirical comparison between $\algname$ and the well-known \emph{triangle-free} process that has recently been shown to produce triangle-free graphs (our problem when $k=3$) with an almost uniform distribution \citep{Morris2013,BohmanKeevash2013}. The comparison shows that the output distribution of $\algname$ is much closer to the uniform distribution.

\subsection{Organization of the Paper}\label{subsec:organization}

The rest of the paper is organized as follows. \S
\ref{sec:related-literature} discusses related research. Description of $\algname$ and the
main result are presented in \S \ref{sec:alg-main-res}. \S \ref{sec:idea} provides the main idea
behind $\algname$ followed by its analysis in \S \ref{sec:analysis}.
An efficient implementation of $\algname$ is presented in \S \ref{sec:run-time} and a comparison with the triangle-free process is given in \S \ref{sec:compare-w-TF}. Finallly, an extension of $\algname$ to bipartite graphs with given degrees is discussed in \S \ref{sec:application}.

%
%

\section{Related Literature}
\label{sec:related-literature}

Random graph models have been used in a wide variety of research areas. For example they are used in determining the effect of having overweight friends in adolescent obesity \citep{Obesity}, in studying social networks that result from uncoordinated random connections created by individuals \citep{JacksonWatts2002}, in modeling emergence of the world wide web as an endogenous phenomena \citep{Papadimitriou2003} with certain topological properties \citep{Kleinberg2000,Newman2003}, and in simulating networking protocols on the Internet topology \citep{inet,faloutsos,medina,bu}.

In information theory, random graphs are used to construct LDPC codes that can approach
Shannon capacity \citep{MCT}, specifically, when the graphs representing the codes are selected uniformly at random
from the set of bipartite graphs with
given degree sequences \citep{AMU06,Chung,LubyEtAl}.
While these random graphs guarantee optimal performances asymptotically,
in practice the LDPC graph has between $10^3$ and
$10^5$ nodes where it is shown
that the existence of a small number of subgraphs
spoil the code
performances \citep{Stopping,Trapping,Pseudo}.
The present paper studies a specific class of such subgraphs (short cycles), but we expect our approach to be applicable to other subgraphs as well.
In addition, for the sake of simplicity, we present the relevant
proofs only for the problem of generating random
graphs without short cycles (not necessarily bipartite nor
with prescribed degrees).
Then we will adapt the algorithm for
generating random bipartite graphs with given degree sequences that have no short cycles\footnote{Implementation details of the application to LDPC codes can be found in this conference paper \citep{ITWpaper}.}. Generalizing proofs to this case is cumbersome but we expect that to be conceptually straightforward.

Random graph generation has also been studied extensively as an important theoretical problem \citep{Wormald1999,Ioannides2006}.
From a theoretical perspective, our work is related to the following
problem. Consider a graph property $P$ that is preserved by removal of
any edge from the graph.
It is a standard problem in extremal graph theory to
determine the largest $m$ such that there exists a graph
with $n$ vertices and $m$ edges having property $P$.
Lower bounds on $m$ can be obtained through the analysis
of greedy algorithms. Such algorithms proceed by sequentially choosing
an edge uniformly from edges whose inclusion would not destroy property $P$, adding that to the graph,
and repeating the procedure until no further edge can be added.
The resulting graph is a random maximal $P$-graph.
The question of finding the
number of edges of a random maximal $P$-graph for several properties
$P$ has attracted considerable attention \citep{RuW92,ESW95,Spe95,BoR00,OsT99,BohmanKeevash,Wolfovitz11,Morris2013,BohmanKeevash2013,Warnke}. In
particular, when $P$ is the property that the graph has no cycles of length $k$, the above process of sequentially growing the graph
is called \emph{$C_k$-free process}.
\cite{BohmanKeevash} showed that the process asymptotically leads to graphs with at least some constant times $n(n\log n)^{1/(k-1)}$ edges which improved earlier results of \cite{BoR00} and \cite{OsT99}. For the case of $k=3$, \cite{Morris2013,BohmanKeevash2013}  proved a sharper result that with high probability (as $n$ goes to $\infty$) the number of edges $m$ would be $[1+o(1)]n\sqrt{n\log(n)/8}$ which is of order $n^{1.5}$ up to logarithmic factors.

In addition to the bound on $m$, and related to the topic of this paper, the analyses by \cite{Morris2013,BohmanKeevash2013} show that certain graph parameters in the $C_3$-free process (also known as triangle-free process) concentrate around their value in uniformly random $C_3$-free graphs. But these papers do not provide any formal statement on closeness of the two distributions. In contrast, we prove that $\algname$ with $k=3$, which is a variant of the $C_3$-free process, generates graphs with a distribution that converges in total-variation distance to uniform $C_3$-free graphs, early in the process; i.e., when $m$ is of order $n^{1+1/36}$. We also provide the rate of this convergence. We note that this range of $m$ is a small subset of the range studied by \citep{Morris2013,BohmanKeevash2013}, but in \S \ref{sec:compare-w-TF} we show that our convergence results are sharper and provide stronger concentration for the graph parameters. In \S \ref{sec:compare-w-TF}, we also emprically demonstrate that the output distribution of $\algname$ is much closer to uniform than the $C_3$-process.

However, we believe the value of $\algname$ and its analysis is when the objective is a more general problem; generating graphs with a given degree sequence that do not have small cycles. In this setting we expect the natural extension of the $C_k$-free process would lead to $C_k$-free graphs with a highly non-uniform distribution. This is motivated by \citep{BKS07} that showed, when the degree sequence is irregular, the process of adding edges uniformly at random in the configuration model, while avoiding creation of double-edges or self-loops, generates graphs with a distribution that is asymptotically equal to the uniform distribution multiplied by an exponentially large bias\footnote{For regular graphs \citep{StW99,Kim-Vu,BKS07} provide a positive result; the output distribution becomes asymptotically uniform when the degrees of are order $\sqrt{n}$.}. However, providing such a rigorous analysis, when the constraint of avoiding small cycles
is added, is still an open problem. We view the present paper as a first step in this direction since it suggests a design approach for the problem (see \S \ref{sec:idea} for details). But to simplify the presentation, we focus the rigorous analysis to the case where the degree sequence constraint is relaxed to just having a fixed number of edges. And in \S \ref{sec:application}, we demonstrate how the approach translates to an algorithm when the degree sequence is prescribed and the graph is bipartite.

This paper is also closely related to the literature on designing sequential algorithms for counting and generating random graphs with given degrees
\citep{ChenDiaconisHolmsLiu,JoePersi,StW99,Kim-Vu,BKS07,blanchet}. In fact, the current paper builds on this line of research and develops two mainly new techniques: (1) for obtaining probabilities $p_{ij}$, instead of starting from a biased algorithm, characterizing its bias, and selecting $p_{ij}$ that can cancel the bias, we use Poisson approximation to directly estimate correct probabilities $p_{ij}$ that leads to an unbiased algorithm, and (2) for the analysis, we use graph approximation methods, Janson inequality, and a combinatorial argument to track the accumulated error from sequentially approximating $p_{ij}$ in each round.

Finally, we note that a preliminary and weaker version of our main result has appeared in proceedings of annual ACM-SIAM Symposium on Discrete Algorithms \citep{SODA_Version}. In particular, Theorem 3.1 of \cite{SODA_Version} only shows that the total variation distance between the output distribution (for a different version) of $\algname$ and the uniform distribution converges to $0$ as size of the graphs goes to $\infty$. But here, we characterize size of the total variation distance for any finite $n$, that is of order $n^{-1/2+k(k+3)\alpha}$. In addition, the aforementioned discussion on $C_k$-free process and its comparison with $\algname$, in \S \ref{sec:compare-w-TF}, are new.

%
%

\section{Algorithm $\algname$ and Main Result}\label{sec:alg-main-res}

In this section we start by introducing some notation and then present our algorithm ($\algname$) followed by
the main theorem on its asymptotic performance.

The \emph{girth} of a graph $G$ is defined to be the length of its shortest cycle.
Let $\gens_{n,m}$ denote the set
of all simple graphs with $m$ edges over $n$ vertices and let
$\gens_{n,m,k}$ be the subset of graphs in $\gens_{n,m}$  with girth greater
than $k$. Throughout the paper $k$ is a constant
and is independent of $n$ and $m$. For any positive integer $s$, the set of integers $1,2,\ldots,s$
is denoted by $[s]$. The complete graph with vertex set $[n]$ is denoted by $K_n$.
For a graph $G$ with $n$ vertices, we label its vertices by integers in $[n]$. For each pair of distinct integers $i,j\in[n]$, an edge that connects node $i$ to node $j$ is denoted by $(ij)$. All graphs considered in this paper are undirected which means $(ij)$ and $(ji)$ refer to the same edge.

$\algname$ starts with an empty graph $G_0$ on $n$ vertices and at each step $t$, $t\in\{0,1,\ldots,m-1\}$, an edge $(ij)$ is added to $G_t$ from $Q(G_t)$,
the set of edges that their addition to $G_t$ does not create a cycle of length at most $k$. Then $G_{t+1}$ will be $G_t\cup(ij)$. If $Q(G_t)$ is the empty set for some $t<m$ then $\algname$ reports ${\sf FAIL}$ and terminates.
The main technical step in $\algname$ is that the edge $(ij)$ is selected randomly from $Q(G_t)$, according to a carefully constructed probability distribution that is denoted by $p(ij|G_t)$ and is given by
\begin{equation}
p(ij|G_t)\equiv \frac{1}{Z(G_t)}e^{-E_k(G_t,ij)}\,.\label{eq:P(G_t)}
\end{equation}
Here $Z(G_t)\equiv\sum_{(ij)\in Q(G_t)}e^{-E_k(G_t,ij)}$ is a normalizing term,
\[
E_k(G_t,ij)\equiv\sum_{r=3}^k\sum_{\ell=0}^{r-2}N_{r,\ell}^{G_t,ij}q_t^{r-1-\ell}\,,
\]
$q_t\equiv\frac{m-t}{{n\choose2}-t}$, and $N_{r,\ell}^{G_t,ij}$ is the number of simple cycles (cycles that do not repeat a vertex) in $K_n$ that have length $r$, include $(ij)$, and include exactly $\ell$ edges of $G_t$. We will provide the intuition behind this complex-looking formula in \S \ref{sec:idea}. In addition, in \S \ref{sec:run-time} we will provide an efficient way of calculating $p(ij|G_t)$ using sparse matrix multiplication.
Throughout the paper, to simplify the notation, in mathematical formula we will refer to $\algname$ by the short notation $\alg$.
\begin{algorithm}
\SingleSpacedXI
\begin{algorithmic}
\State \textbf{Input:} $n$, $m$, $k$
\State \textbf{Output:} An element of $\gens_{n,m,k}$ or {\sf FAIL}
\State set $G_0$ to be a graph over vertex set $[n]$ and with no edges
\For {each $t$ in $\{0,\ldots,m-1\}$}
\If {$|Q(G_t)|=0$}
\State stop and return {\sf FAIL}
\Else
\State sample an edge $(ij)$ with probability $p(ij|G_t)$, defined by Eq. \eqref{eq:P(G_t)}
\State set $G_{t+1}= G_t\cup (ij)$
\EndIf
\EndFor
\If {the algorithm does not {\sf FAIL} before $t=m-1$}
\State return $G_m$
\EndIf
\end{algorithmic}
\caption{$\algname$.}
\end{algorithm}

By construction, if $\algname$ outputs a graph $G$ then $G$ is a member of $\gens_{n,m,k}$. If $\algname$ outputs {\sf FAIL} the algorithm will be repeated till it produces a graph. We will show later that the probability of {\sf FAIL} output vanishes asymptotically.
Let $\P_{\alg}(G)$ be the
probability that $\algname$ does not {\sf FAIL} and returns graph $G$.
Let also $\P_{\unif}$ be the uniform
probability on the set $\gens_{n,m,k}$; that is
$\P_{\unif}(G)=1/|\gens_{n,m,k}|$. Our goal is to show that $\P_{\alg}(G)$ and $\P_{\unif}(G)$ are very close in total variation distance.
The \emph{total variation distance} between two probability measures
$\P$ and $\mb{Q}$ on a set $X$ is defined by $\dv(\P,\mb{Q})\equiv \sup\,\Big\{|\P(A)-\mb{Q}(A)|\,:\,A\subset X\Big\}.$ Now, we are ready to state the main result of the paper. Its proof is provided in \S \ref{sec:analysis}.
\begin{theorem}\label{thm:main}
For $m=O(n^{1+\alpha})$, $m\ge n$, and a constant $k\geq 3$ such that
$\alpha\leq 1/[2k(k+3)]$, the failure probability of $\algname$ asymptotically vanishes and
the graphs generated by $\algname$ are approximately uniform. In particular,
\[
\P_{\alg}({\sf FAIL})=O(n^{-1/2+k(k+3)\alpha})~~~\textrm{and}~~~\dv(\P_{\alg},\P_{\unif})=O(n^{-1/2+k(k+3)\alpha})\,.
\]%
\end{theorem}
The next result shows a run-time guarantee for $\algname$ and is proved in \S \ref{sec:run-time}.
\begin{theorem}\label{thm:runtime}
Let $n$, $m$, and $k$ satisfy the conditions of Theorem \ref{thm:main}. For all $n$ large enough, there exist an implementation of
$\algname$ that uses asymptotically $O(n^2m)$ operations in expectation.
\end{theorem}

%
%

\section{The Intuition Behind $\algname$}\label{sec:idea}

In order to understand $\algname$, and in particular the calculations for $[p(ij|G_t)]$, it is instructive to examine the
 \emph{execution tree}
$\mrm{T}$ of a simpler version of $\algname$ that sequentially adds $m$ random edges to an empty graph on $n$ vertices to obtain an element of $\gens_{n,m}$ (without
any attention to whether a short cycle is generated or not).
Consider a rooted $m$-level tree where the root (the vertex in level
zero) corresponds to the empty graph at the beginning of this sequential algorithm and level $t$ vertices correspond to all
pairs $(G_t,\pi_t)$ where $G_t$ is a partial graph
that can be constructed after $t$ steps, and $\pi_t$ is an ordering of its $t$
edges. There is a link (edge) in
$\mrm{T}$ between a partial graph $(G_t,\pi_t)$ from level $t$ to a
partial graph $(G_{t+1},\pi_{t+1})$ from level $t+1$ if
$G_t\subset G_{t+1}$ and the first $t$ edges of $\pi_t$ and $\pi_{t+1}$ are equal.
Any path from the root to a leaf at level $m$ of
$\mrm{T}$ corresponds to one possible way of sequentially generating a random
graph in $\gens_{n,m}$.

Let us denote those partial graphs $G_t$ that have girth greater than $k$ by \emph{valid} graphs.
Our goal is to reach a valid leaf in
$\mrm{T}$, uniformly at random, by starting from the root and going down the tree. A myopic approach could be repeating the above sequential algorithm
many times until its output in step $m$ is a valid leaf of $\mrm{T}$. However,
when $m =O(n^{1+\alpha})$, the fraction of valid leaves
is of order $e^{-n^{\alpha}}$ (see \S \ref{sec:analysis} for details). Therefore, this myopic approach has an exponentially small chance of success. Note that the myopic approach works well when $m=O(n)$ since a constant fraction of leaves of $\mrm{T}$ are valid. Therefore, our focus is when $m$ is super linear in $n$.

In contrast to the myopic approach, $\algname$ is designed based on a general strategy for uniformly randomly generating valid leaves of $\mrm{T}$ \citep{SinclairBook};
at any step $t$, it chooses $(ij)$ with probability proportional to the number of valid leaves of $\mrm{T}$ among descendant of $(G_{t+1},\pi_{t+1})$ where $G_{t+1}=G_t\cup (ij)$. Denote this probability by $p_{\true}(G_{t+1},\pi_{t+1})$. The main challenge for implementing this strategy is calculating $p_{\true}(G_{t+1},\pi_{t+1})$. In $\algname$
we will approximate $p_{\true}(G_{t+1},\pi_{t+1})$  with  $p(G_{t+1},\pi_{t+1})$ as follows. Let $n_k(G_{t+1},\pi_{t+1})$ denote the number of cycles of length at most $k$ in a leaf chosen uniformly at random among descendants of  $(G_{t+1},\pi_{t+1})$ in $\mrm{T}$. Note that $p_{\true}(G_{t+1},\pi_{t+1})$ is by definition equal to
$\mathbb{P}\left\{n_k(G_{t+1},\pi_{t+1})=0\right\}$. Using Poisson approximation, see \citep{AlS92} for details, one expects the distribution of $n_k(G_{t+1},\pi_{t+1})$ to be approximately Poisson. In particular,
\begin{equation}\label{eq:nk-poisson}
\P\{n_k(G_{t+1},\pi_{t+1})=0\}\approx\exp\left(-\mb{E }[n_k(G_{t+1},\pi_{t+1})]\right)\,.
 \end{equation}
Therefore, our approximation $p(G_{t+1},\pi_{t+1})$ will be chosen to be proportional to the right hand side of Eq. \eqref{eq:nk-poisson}. This is the main intuition behind Eq. \eqref{eq:P(G_t)}.

A crucial step in the analysis of $\algname$, provided in \S \ref{sec:analysis}, is to control the \emph{accumulated error}
\[
\prod_{t=0}^{m-1} \left[\frac{p(G_{t+1},\pi_{t+1})}{p_{\true}(G_{t+1},\pi_{t+1})}\right]\,.
\]
Prior work \citep{Kim-Vu,BKS07} used sharp concentration inequalities to find a separate upper bound, for each $t$, on the error term
$\left[p(G_{t+1},\pi_{t+1})/p_{\true}(G_{t+1},\pi_{t+1})\right]$. Instead, in this paper we
simplify the final product $\prod_{t=0}^{m-1} \left[p(G_{t+1},\pi_{t+1})/p_{\true}(G_{t+1},\pi_{t+1})\right]$ and will approximate it directly which leads to a tighter bound.

%
%

\section{Analysis of $\algname$ and Proof of Theorem \ref{thm:main}}\label{sec:analysis}

The aim of this section is to prove Theorem \ref{thm:main}.
The core of the proof is to show that $\P_{\alg}(G)$, probability of generating a graph $G$ by $\algname$, is asymptotically larger than $\P_{\unif}(G)$, the uniform probability over $\gens_{n,m,k}$. After this result is stated in Lemma \ref{lem:mainlowerbound}, it is used to prove Theorem \ref{thm:main}. The rest of the section is divided to four subsections. In particular, \S \ref{ssec:lbound-4-PX.} describes the main steps for proving Lemma \ref{lem:mainlowerbound} which rely on auxiliary Lemmas \ref{lem:numGraphs} and \ref{lem:S1-S2-S3-lbd}. These auxiliary lemmas are stated in \S \ref{ssec:lbound-4-PX.} and proved in \S \ref{sec:pf-lem-numGraphs} and \S \ref{sec:pf-lem-Si-lbd} respectively. Throughout this section we will introduce a large number of new notations. For convenience, we have repeated all notations with their definition in Table \ref{tab:notations} of Appendix \ref{app:notations}.

\begin{lemma}\label{lem:mainlowerbound}
There exist positive constants $c_1$ and $c_2$ such that
\[
\P_{\alg}(G) \geq \left[1-c_1n^{-1/2+k(k+3)\alpha}\right]\P_{\unif}(G)\,,
\]
for every $n,m,k$ satisfying the conditions of Theorem \ref{thm:main}, and all $G\in\gens_{n,m,k}$ except for a subset of graphs in $\gens_{n,m,k}$ of size $c_2\exp(-n^{k\alpha})|\gens_{n,m,k}|$.
\end{lemma}
In other words, Lemma \ref{lem:mainlowerbound} shows that for all but $o(|\gens_{n,m,k}|)$ graphs $G$ in $\gens_{n,m,k}$
inequality
$
\P_{\alg}(G)\geq [1-o(1)]\P_{\unif}(G)
$,
holds where the term $o(1)$ goes to zero as $n$ goes to infinity uniformly in the graph $G$. Next, we prove Theorem \ref{thm:main} using Lemma \ref{lem:mainlowerbound}.
\proof{Proof of Theorem \ref{thm:main}}
From the definition of $\dv(\P_{\alg},\P_{\unif})$, using triangle inequality, we obtain
\[
\dv(\P_{\alg},\P_{\unif})\leq \sum_{G\in\gens_{n,m,k}}|\P_{\alg}(G)-\P_{\unif}(G)|\,.
\]
Then, depending on whether $\P_{\alg}(G)\geq \P_{\unif}(G)$ or $ \P_{\alg}(G)<[1-c_1n^{-1/2+k(k+3)\alpha}]\P_{\unif}(G)$ we bound the term $|\P_{\alg}(G)-\P_{\unif}(G)|$ differently.
Let $\sB_{n,m,k}\subset\gens_{n,m,k}$ be the set of all graphs $G$ with $\P_{\alg}(G)\leq\P_{\unif}(G)$ and
let the subset $\sD_{n,m,k}\subseteq\sB_{n,m,k}$ to be those graphs $G$ in $\sB_{n,m,k}$ with
$\P_{\alg}(G)<[1-c_1n^{-1/2+k(k+3)\alpha}]\P_{\unif}(G)$. To simplify the notation, for the rest of the proof we drop the subscripts $n,m,k$ from
$\sB_{n,m,k},\sD_{n,m,k}$ and $\gens_{n,m,k}$. Assuming Lemma \ref{lem:mainlowerbound} holds then
$|\sD|=c_2\,e^{-n^{k\alpha}}\,|\gens|$ and for $G\in \sB\backslash \sD$
\begin{align}
|\P_{\alg}(G)-\P_{\unif}(G)|=\P_{\unif}(G)-\P_{\alg}(G)&\leq c_1n^{-1/2+k(k+3)\alpha}\,\P_{\unif}(G) \label{eq:|Palg-Punif|-UpperBound} \,.
\end{align}
Therefore,
\begin{align}
\sum_{G\in\gens}\Big|\P_{\alg}(G)-\P_{\unif}(G)\Big|&=
\sum_{G\in\gens}\Big[\P_{\alg}(G)-\P_{\unif}(G)\Big] + 2\sum_{G\in \sB}\Big|\P_{\alg}(G)-\P_{\unif}(G)\Big|\label{eq:sum|PX-PU|<o(1)}\\
&=\sum_{G\in\gens}\Big[\P_{\alg}(G)-\P_{\unif}(G)\Big]+ 2\sum_{G\in \sB\backslash \sD}\Big|\P_{\alg}(G)-\P_{\unif}(G)\Big|
+2\sum_{G\in  \sD}\Big|\P_{\alg}(G)-\P_{\unif}(G)\Big|\no\\
&\stackrel{(a)}{\leq}\sum_{G\in\gens}\P_{\alg}(G)-\sum_{G\in\gens}\P_{\unif}(G)+2c_1n^{-1/2+k(k+3)\alpha}\,\sum_{G\in \sB\backslash\sD}\P_{\unif}(G)+ 4\sum_{G\in \sD}\P_{\unif}(G)\no\\
&\leq 1-\P_{\alg}({\sf FAIL})-1 + 2c_1n^{-1/2+k(k+3)\alpha}+4\frac{|\sD|}{|\gens|}\no\\
&\leq 2c_1n^{-1/2+k(k+3)\alpha}+4c_2e^{-n^{k\alpha}}-\P_{\alg}({\sf FAIL})\,,\no
\end{align}
where $(a)$ uses Eq. \eqref{eq:|Palg-Punif|-UpperBound} and triangle inequality. Also, $\P_{\alg}({\sf FAIL})$ is the probability of failure of $\algname$.  In summary, we proved
\begin{align}
\dv(\P_{\alg},\P_{\unif})+ \P_{\alg}({\sf FAIL})\leq \sum_{G\in\gens}|\P_{\alg}(G)-\P_{\unif}(G)|+ \P_{\alg}({\sf FAIL})=O(n^{-1/2+k(k+3)\alpha})\,,\no
\end{align}
which finishes the proof $\square$
\endproof

Throughout the rest of this section our focus will be on proving Lemma \ref{lem:mainlowerbound}.

%
%
\subsection{Lower Bound For $\P_{\alg}(G)$: Proof of Lemma \ref{lem:mainlowerbound}}\label{ssec:lbound-4-PX.}

We break proof of Lemma \ref{lem:mainlowerbound} into four main steps. Two of these steps (steps 1 and 3 below) will be major and involve proving additional Lemmas that will be later proved in \S \ref{sec:pf-lem-numGraphs} and \S \ref{sec:pf-lem-Si-lbd}.
%
%
\proof{Step 1 in Proof of Lemma \ref{lem:mainlowerbound}: Approximating $\P_{\unif}$ via Jansen inequality.} Since $\P_{\unif}=1/|\gens_{n,m,k}|$, we will find an asymptotic estimate for $|\gens_{n,m,k}|$ using Janson inequality \citep{Janson90} that shows the number of cycles of constant
length in $\gens_{n,m}$ is approximately a Poisson random variable. The result is summarized in the following lemma that is proved in \S \ref{sec:pf-lem-numGraphs}. Before stating the lemma, we define $\C_r$ to be the set of all simple cycles of length $r$ in $K_n$ and introduce notation $N$ for total number of edges in $K_n$ which is equal to ${n\choose2}$.
\begin{lemma}\label{lem:numGraphs}
Let $m=O(n^{1+\alpha})$ with $\alpha<1/(2k-1)$, $k\geq 3$, and $m\ge n$, then
\begin{align}
\frac{\P_{\unif}(G)}{\left\{{N\choose m}\exp\le[-\sum_{r=3}^{k}|\C_r|\le(\frac{m}{N}\ri)^r\ri]\right\}^{-1}}=e^{O\left(n^{\frac{3k\alpha-1}{2}}\right)}\,.
\label{eq:asympt-4-numgraphs}
\end{align}
In other words, the number of graphs with $n$ vertices, $m$ edges, and no cycle of length up to $k$ is $(1+o(1)){N\choose m}\exp[-\sum_{r=3}^{k}|\C_r|(m/N)^r]$ where the $o(1)$ term is of order $n^{\frac{3k\alpha-1}{2}}$.
\end{lemma}
The remaining steps will provide necessary approximations and algebraic simplifications to find an asymptotic lower bound for $\P_{\alg}$ which will be equal to
the denominator term in Eq. \eqref{eq:asympt-4-numgraphs}.

\proof{Step 2 in Proof of Lemma \ref{lem:mainlowerbound}: Using convexity and Jensen Inequality.} Let us start by writing an expression for $\P_{\alg}(G)$ when $G$ is a \emph{fixed} element of $\gens_{n,m,k}$.  Note that $\algname$ sequentially adds edges to an empty graph to produce a graph with $m$ edges. Hence for the fixed graph $G$, there are $m!$ permutations of the edges of $G$ that can be generated by  $\algname$ and each permutation can be output with a different probability.  Let $\pi$ be any permutation of edges of $G$ (i.e. a one-to-one mapping from
$\{1,\dots,m\}$ to the edges of $G$), and let $G^{\pi}_t$ be the graph
having $[n]$ as vertex set and $\{\pi(1),\dots,\
\pi(t)\}$ as edge set. This is the partial graph that is generated after $t$ steps of $\algname$ conditioned on having $\pi$ as output.  Now we can write
\[
\P_{\alg}(G)  =  \sum_{\pi}\prod_{t=0}^{m-1}p(\pi(t+1)|G^{\pi}_t)\,.
\]
Additionally, consider the uniform distribution on the set of all $m!$ permutations $\pi$. Then, $\sum_{\pi}$ can be replaced by $m!\,\E_\pi$ where $\E_\pi$ is expectation with respect to a random permutation $\pi$. Hence,
\begin{align}
\P_{\alg}(G)= m!\,\E_{\pi}\le\{\prod_{t=0}^{m-1}p(\pi(t+1)|G^{\pi}_t)\ri\}
&=m!\, \E_{\pi}\exp\le\{\sum_{t=0}^{m-1}\log p(\pi(t+1)|G^{\pi}_t)\ri\}\no\\
&\geq m!\,\exp\le\{\sum_{t=0}^{m-1}\E_{\pi}\log p(\pi(t+1)|G^{\pi}_t)\ri\}\,,\label{eq:PG-Jensen}
\end{align}
where the inequality is by Jensen inequality for the convex function $e^x$. Next, applying the definition of $p(\pi(t+1)|G_t)$ from Eq. \eqref{eq:P(G_t)} we get
\begin{align}\label{eq:LB-1}
\P_{\alg}(G)\ge m!\, \exp \Bigg[-\sum_{t=0}^{m-1}\E_{\pi}~E_k(G_t^\pi,\pi(t+1))-\sum_{t=0}^{m-1}\E_{\pi}\log Z(G_t^{\pi})\Bigg]\,.
\end{align}
Now, we define $F(G_t^\pi)$ to be the set of all \emph{forbidden} pairs at step $t$, pairs of nodes $i$ and $j$ that adding $(ij)$ to $G_t^\pi$ creates a cycle of length at most $k$, and set $Z_0(G_t^\pi)\equiv N-t-|F(G_t^\pi)|$. Note that,
\begin{eqnarray}
\log Z(G_t^\pi)&=&\log Z_0(G_t^\pi)+\log\frac{Z(G_t^\pi)}{Z_0(G_t^\pi)}\no\\
&=&\log \le[(N-t)(1-\frac{|F(G_t^\pi)|}{N-t})\ri]+\log\frac{Z(G_t^\pi)}{Z_0(G_t^\pi)}\no\\
&\leq&\log(N-t)-\frac{|F(G_t^\pi)|}{N-t}+\log\frac{Z(G_t^\pi)}{Z_0(G_t^\pi)}\,,\label{eq:LB-2}
\end{eqnarray}
using inequality $\log(1-x)\leq -x$ for $x\in (-\infty,1]$ that holds since $|F(G_t^\pi)|\leq N-t$. Combining
Eqs. \eqref{eq:LB-1} and \eqref{eq:LB-2} and using $1/(N-t)\ge 1/N$, we arrive at the following modified lower bound for $\P_{\alg}(G)$
\begin{align}
\P_{\alg}(G) &\geq \frac{1}{{N\choose m}}\,\exp \left\{
\underbrace{\left[-\sum_{t=0}^{m-1}\E_\pi E_k(G_t^\pi,\pi(t+1))\right]}_{S_1(G)}
+\underbrace{\left[\frac{1}{N}\sum_{t=0}^{m-1} \E_{\pi} |F(G_t^\pi)|\right]}_{S_2(G)}
+\underbrace{\left[-\sum_{t=0}^{m-1}\E_{\pi}\log \frac{Z(G_t^{\pi})}{Z_0(G_t^{\pi})}\right]}_{S_3(G)}
\right\}\,.\label{eqn:1}
\end{align}
The next step is the most important part of our effort in the journey to prove Lemma \ref{lem:mainlowerbound}.

\proof{Step 3 in Proof of Lemma \ref{lem:mainlowerbound}: Simplifying $S_1(G)+S_2(G)+S_3(G)$.} This step shows the main benefit of deferring the calculation of approximation errors for $p(ij|G_t^\pi)$ to the final step.  We will show that even though the terms $S_i(G)$ for $i=1,2,3$ can be large and dependent on $G$, many terms in their combined sum cancel out and the resulting expression will be independent of $G$. In particular, we will show that the only negative term\footnote{$S_3(G)$ will be positive since $Z(G_t^\pi)<Z_0(G_t^\pi)$.}, $S_1(G)$, will completely cancel $S_2(G)$ and all graph dependent parts of $S_3(G)$. Throughout the rest, since $G$ is fixed, we often drop the references to $G$ in $S_i:\,i=1,2,3$.

The main result of this step is summarized in the following lemma. First we define
$\C_{r,\ell}(G)$ to be the set of all simple cycles of length $r$, belonging to $K_n$, that include exactly $\ell$ edges of $G$.
\begin{lemma}\label{lem:S1-S2-S3-lbd} Let $m$ be larger than $n$ and also satisfy $m=O(n^{1+\alpha})$ where $\alpha\leq1/[2k(k+3)]$ for a constant $k\geq 3$. Then for all but $O(e^{-n^{k\alpha}})$ fraction of graphs $G$ in $\gens_{n,m,k}$ the three inequalities below hold. In other words, the number of graphs in $\gens_{n,m,k}$ that violate at least one of the inequalities has size of order $e^{-n^{k\alpha}}|\gens_{n,m,k}|$.
\begin{itemize}
\item[(a)] $S_1(G)\geq -O\left(n^{(k-1)(k+3)\alpha-1}\right)-\sum_{r=3}^k\sum_{\ell=1}^{r-1}|\C_{r,\ell}(G)|\le(\frac{m}{N}\ri)^{r-\ell}\ell\int_0^1\theta^{\ell-1}(1-\theta)^{r-\ell}d\theta.$
\vspace{2mm}

\item[(b)]$S_2(G)\geq -O\Big(n^{k(k+3)\alpha-1/2}\Big)+\sum_{r=3}^k|\C_{r,r-1}(G)|\le(\frac{m}{N}\ri)\int_0^1\theta^{r-1}d\theta$.
\vspace{2mm}

\item[(c)]$S_3(G)\geq -O(n^{k(k+3)\alpha-1/2})+\sum_{r=3}^k\sum_{\ell=0}^{r-2}|\C_{r,\ell}(G)|(\frac{m}{N})^{r-\ell}(r-\ell)\int_{0}^{1}\theta^{\ell}(1-\theta)^{r-\ell-1}d\theta$.
\vspace{2mm}

\end{itemize}
\end{lemma}
We defer proof of Lemma \ref{lem:S1-S2-S3-lbd} to \S \ref{sec:pf-lem-Si-lbd}.

\proof{Step 4 and the Final Step in Proof of Lemma \ref{lem:mainlowerbound}.}

Next we will show how the different terms in lower bounds for $S_i$'s from Lemma \ref{lem:S1-S2-S3-lbd} cancel each other.
The main idea in relating the terms in the lower bounds is the following equation which is obtained using integration by parts for $r-1\geq\ell>1$,
\begin{equation}
\label{eqn:int-by-parts}
\ell\int_0^1\theta^{\ell-1}(1-\theta)^{r-\ell}d\theta=(r-\ell)\int_{0}^{1}\theta^{\ell}(1-\theta)^{r-\ell-1}d\theta\,.
\end{equation}
Using \eqref{eqn:int-by-parts} we can see that, when adding the right hand sides of the three inequalities in Lemma \ref{lem:S1-S2-S3-lbd}, all terms in the lower bound for $S_1$ with $1\leq\ell\leq r-2$ are canceled with the corresponding terms in the lower bound for $S_3$. In addition, the $\ell=r-1$ terms in the lower bound of $S_1$ are canceled with the lower bound of $S_2$. Therefore, the uncanceled terms are $\ell=0$ terms from the lower bound of $S_3$ which we will see below to be asymptotically independent of $G$. More formally, combining Eq. \eqref{eqn:1} and Lemma \ref{lem:S1-S2-S3-lbd}, for all graphs $G$ in $\gens_{n,m,k}$ except a subset of size $O(e^{-n^{k\alpha}}|\gens_{n,m,k}|)$,
\begin{eqnarray}
\P_{\alg}(G)&\geq& \frac{1}{{N\choose m}} \exp\left[S_1(G)+S_2(G)+S_3(G)\right]\,\no\\
&\geq& \frac{1}{{N\choose m}}\exp\left[-O(n^{k(k+3)\alpha-1/2})+\sum_{r=3}^k|\C_{r,0}(G)|\le(\frac{m}{N}\ri)^r~r\int_{0}^{1}(1-\theta)^{r-1}d\theta\right]\,\no\\
&=& \frac{1}{{N\choose m}}\exp\left[-O(n^{k(k+3)\alpha-1/2})+\sum_{r=3}^k|\C_{r,0}(G)|\le(\frac{m}{N}\ri)^r\right]\,.\label{eq:PS>=e^(S1+S2+S3)}
\end{eqnarray}
We note that even though the equality \eqref{eqn:int-by-parts} is just an algebraic fact, it can be viewed as double-counting a combinatorial quantity using two different approaches. The quantity would be number of times a cycle in $K_n$ would be considered in calculation of probability terms $p(\pi(t+1)|G_t^\pi)$. In \S \ref{sec:pf-lem-Si-lbd} we perform both counting arguments and then approximate the result of each counting argument with integration with respect to $\theta=t/m$.

Comparing \eqref{eq:PS>=e^(S1+S2+S3)} and the asymptotic expression for $\P_{\unif}(G)$ given by the denominator in left hand side of Eq. \eqref{eq:asympt-4-numgraphs}, we see that the only difference in the exponent is the use of $|\C_{r,0}(G)|$ instead of $|\C_{r}|$ and the following lemma, proved in \S \ref{app:auxillary}, provides the final piece.
\begin{lemma}\label{lem:Cr0/Cr} If $m=O(n^{1+\alpha})$ and $k$ is constant then $|\C_r\backslash\C_{r,0}(G)|/|\C_r|=O(n^{\alpha-1})$.
\end{lemma}
Using Lemma \ref{lem:Cr0/Cr} we have
\begin{eqnarray*}
\sum_{r=3}^k|\C_{r,0}(G)|\le(\frac{m}{N}\ri)^r\geq
\sum_{r=3}^k|\C_r|\,\left[1-O(n^{\alpha-1})\right]\le(\frac{m}{N}\ri)^r
\geq-O(n^{(k+1)\alpha-1})+\sum_{r=3}^k|\C_r|\le(\frac{m}{N}\ri)^r\,,
\end{eqnarray*}
where the last inequality uses $|\C_r|=O(n^r)$ and $m=O(n^{1+\alpha})$. Summarizing, using Lemmas \ref{lem:numGraphs}-\ref{lem:S1-S2-S3-lbd}, for all graphs $G$ in $\gens_{n,m,k}$ except a subset of size
$O(e^{-n^{k\alpha}}|\gens_{n,m,k}|)$ we have
\begin{align*}
\P_{\alg}(G)
&\geq\frac{\exp\left[-O(n^{k(k+3)\alpha-1/2})+\sum_{r=3}^k|\C_r|\le(\frac{m}{N}\ri)^r\right]}{{N\choose m}}\\
&\geq \exp\left[-O(n^{k(k+3)\alpha-1/2})-O(n^{(3k\alpha-1)/{2}})\right]\,\P_{\unif}(G)\\
&=\exp\left[-O(n^{k(k+3)\alpha-1/2})\right]\,\P_{\unif}(G)\\
&\ge \left[1-O(n^{k(k+3)\alpha-1/2})\right]\,\P_{\unif}(G)
\,.
\end{align*}
Here the last inequality uses $e^x\ge 1+x$. The above equation means that there is a constant $c_1$ where $\P_{\alg}(G)\ge [1-c_1n^{k(k+3)\alpha-1/2})]\P_{\unif}(G)$ for the same family of graphs which finishes proof of Lemma \ref{lem:mainlowerbound}. Therefore, all we need now is proving Lemmas \ref{lem:numGraphs}-\ref{lem:S1-S2-S3-lbd} $\square$

\endproof
%
%
\subsection{Approximating $|\gens_{n,m,k}|$ and Proof of Lemma \ref{lem:numGraphs}}\label{sec:pf-lem-numGraphs}
\label{ssec:est-Gnmk}

Before delving into the details, we provide a high-level overview of the proof. The main idea is
to look at the random graph model $\gens_{n,m}$ and estimate the probability of the event of having a graph
with girth larger than $k$ using Janson inequality. However, we will do all of this on an approximation to
the random graph model $\gens_{n,m}$, namely random graph model $\gens_{n,p}$
where each edge on vertices of $[n]$ appears independently randomly with probability $p=m/N$.
This type of approximation is well-known in random graph literature \citep{JLR00}.
Any graph in $\gens_{n,p}$ would
have on average $m$ edges, making $\gens_{n,p}$ a natural approximation to $\gens_{n,m}$.

\subsubsection{Approximating $\P_{n,p}(A_k)$ via Janson Inequality.}\label{sssec:Pk-via-Janson} First we define Janson inequality.
\begin{definition}[Janson Inequality]\label{def:Janson-ineq}
Let $\mathbb{I}$ be a set of graphs on the vertex set $[n]$. Now consider a random graph $G$ from $\gens_{n,p}$, for any $i\in\mathbb{I}$ we define a ``bad event'' $B_i$ to be when $G$ contains $i$ as a subgraph. Janson inequality aims to estimate the probability that $G$ does not contain any subgraph in $\mathbb{I}$, that is equal to $\P\big(\cap_{i\in\ind}B_i^{(c)}\big)$, when the
events $\{B_i^{(c)}\}_{i\in\ind}$ are \emph{almost independent}. More formally,
let $\eta,\,\xi$ be real numbers such that and for all $i$ in $\ind$,
\[
\P(B_i)\leq \eta <1 ~~~ \textrm{and}~~~\sum_{B_j\sim B_i}\P(B_i\cap B_j) = \xi\,.
\]
Here $B_i\sim B_j$ means that $B_i$, $B_j$ are dependent which means the subgraphs $i$ and $j$ have at least one common edge.
Then Janson inequality is
\begin{equation}
\label{eq:janson-ineq}
\prod_{i\in\ind}\P(B_i^{(c)})\leq\P\big(\cap_{i\in\ind}B_i^{(c)}\big)\leq\exp\left(\frac{\xi}{2(1-\eta)}\right)\prod_{i\in\ind}\P(B_i^{(c)})\, .
\end{equation}
In particular, for $\xi=o(1)$ we have
$
\P\big(\cap_{i\in\ind}B_i^{(c)}\big)=(1+o(1))\prod_{i\in\ind}\P(B_i^{(c)})$.
\end{definition}
\begin{remark}
Janson inequality is not necessarily about subgraphs of a random graph and is more general. For brevity we stated the inequality in the above form and defer the reader to \citep{Janson90} or \citep{AlS92} for the more general version.
\end{remark}

Let us denote the
probability with respect to the randomness in $\gens_{n,p}$ and
$\gens_{n,m}$ by $\P_{n,p}$ and $\P_{n,m}$ respectively. Let $A_k$
be the event that a random graph, selected from $\gens(n,p)$ or $\gens(m,n)$, has girth greater than $k$. Our next
step is to calculate $\P_{n,p}(A_k)$.

For every cycle $i$
of length at most $k$ on vertices of $[n]$ we consider a bad event $B_i$ that
is the event that a random graph $G$ from $\gens_{n,p}$ contains
cycle $i$. In particular, $\ind = \cup_{r=3}^k\C_r$. It is not difficult to see
that $\P(B_i)= O(p^k)$ and $\xi = O(\sum_{r_1=3}^k\sum_{r_2=3}^kn^{r_1+r_2-2}p^{r_1+r_2-1})$.
And since $p=O(n^{\alpha-1})$ then using Janson inequality
\eqref{eq:janson-ineq},
\begin{equation*}
\prod_{i\in\mathbb{I}}\P(B_i^{(c)})\leq\P_{n,p}(A_k)\leq
e^{O\le(n^{(2k-1)\alpha-1}\ri)}~\prod_{i\in\ind}\P(B_i^{(c)})
\end{equation*}
which gives the following for $\alpha<1/(2k-1)$,
\begin{eqnarray}
\P_{n,p}(A_k)&=&e^{O\le(n^{(2k-1)\alpha-1}\ri)}~~\prod_{i\in\ind}\P(B_i^{(c)})\no\\
&=&e^{O\le(n^{(2k-1)\alpha-1}\ri)}~~\prod_{i\in\ind}\le(1-p^{\textrm{length}(i)}\ri)\no\\
&=&\exp\le[O\le(n^{(2k-1)\alpha-1}\ri)+\sum_{r=3}^k|\C_r|\log(1-p^r)\ri]\no\\
&=&\exp\le[O\le(n^{(2k-1)\alpha-1}\ri)-\sum_{r=3}^k|\C_r|p^r\ri]\,.\label{eq:Pk-est}
\end{eqnarray}
The last step uses $\log(1-x)=-x+O(x^2)$ and
$|\C_r|p^{2r}=O(n^{2r\alpha-r})=O(n^{(2k-1)\alpha-1})$.

\subsubsection{Approximating $\P_{n,m}(A_k)$ with $\P_{n,p}(A_k)$.}

We start by stating the following result on monotone properties of $\gens_{n,p}$ and $\gens_{n,m}$. However, we only state it for the specific event $A_k$ but it applies to more general events that satisfy the following property. If $G$ is in $A_k$ then any graph $G'$, obtained by removal of an edge from $G$, would also be contained in $A_k$. Such events
are known as \emph{monotone graph properties}.
\begin{proposition}[Lemma 1.10 in \citep{JLR00}]
For  $0\leq p\leq p'\leq 1$ and $0\leq m\leq m'\leq N$ we have
$\P_{n,p}(A_k)\geq \P_{n,p'}(A_k)$ and $\P_{n,m}(A_k) \geq \P_{n,m'}(A_k)$.
\end{proposition}
\proof{Proof of Lemma \ref{lem:numGraphs}.} First define $m(G)$ to be the number of edges for any graph $G$. Now we state the following lemma for comparing $\P_{n,p}(A_k)$ and $\P_{n,m}(A_k)$ that is proved in Appendix \ref{app:auxillary}.
\begin{lemma}\label{lem:monotonicity_inequalities}
For any $0<p<1$, $1<m<N$, and the monotone event $A_k$ described above we have
\begin{align}
\P_{n,p}(A_k)&\leq \P_{n,m}(A_k) + \P_{n,p}\Big(\nedge(G)<m\Big)\,,\label{eq:Pq<Pm}\\
\P_{n,p}(A_k)&\geq\P_{n,m}(A_k)-\P_{n,p}\Big(\nedge(G)> m\Big)\,.\label{eq:Pq>Pm}
\end{align}
\end{lemma}

Next, we state a lemma, proved in \S \ref{app:auxillary} using Hoeffding inequality, that provides a sharp upper bound for the probability of the event that a graph $G$ in $\gens_{n,p}$ does not have exactly $m$ edges when $p$ is close to $m/N$.
\begin{lemma}\label{lem:hoeffding-cor}
For $\beta$ with $0<\beta<1$ if $m$ is large enough and $p_1\equiv\frac{m-m^{\frac{1+\beta}{2}}}{N}$ and $p_2\equiv\frac{m+m^{\frac{1+\beta}{2}}}{N}$, we have
\begin{eqnarray}
\P_{n,p_1}\Big(\,\nedge(G)>m\,\Big)&\leq&e^{-m^\beta/8}\,,\label{eq:Hoeffding4p1}\\
\P_{n,p_2}\Big(\,\nedge(G)<m\,\Big)&\leq&e^{-m^\beta/8}\,.\label{eq:Hoeffding4p2}
\end{eqnarray}
\end{lemma}
Now we can use \eqref{eq:Pq>Pm} for $m$ and $p_1$
together with \eqref{eq:Hoeffding4p1} to obtain
\begin{align}
\P_{n,m}(A_k)\leq\P_{n,p_1}(A_k)+\P_{n,p_1}\Big(\nedge(G)>m\Big)
\leq \P_{n,p_1}(A_k)+e^{-\frac{m^\beta}{8}}\,. \label{eq:ineq-1}
\end{align}
Similarly, \eqref{eq:Pq<Pm} for $m$ and $p_2$ combined with \eqref{eq:Hoeffding4p2} gives
\begin{align}
\P_{n,m}(A_k)\geq\P_{n,p_2}(A_k)-\P_{n,p_2}\Big(\nedge(G)<m\Big)
\geq \P_{n,p_2}(A_k)-e^{-\frac{m^\beta}{8}}\,.\label{eq:ineq-2}
\end{align}

\subsubsection{Finalizing Proof of Lemma \ref{lem:numGraphs}.}\label{subsec:proof-lemma-2} First, to simplify the formulas we introduce  new notation that will only be used in \S \ref{subsec:proof-lemma-2} . Recall from \eqref{eq:Pk-est} that $\P_{n,p}=\exp[-H(p)+O(n^{(2k-1)\alpha-1})]$ where $H(p)=\sum_{r=3}^k|\C_r|p^r$. Combining \eqref{eq:ineq-1} and \eqref{eq:ineq-2} and using this new notation we have,
\begin{equation}
\label{eq:sandwich-1}
 e^{H(p)-H(p_2)+O(n^{(2k-1)\alpha-1})}-e^{-\frac{m^\beta}{8}+H(p)} \leq \frac{\P_{n,m}(A_k)}{\exp[-H(p)]} \leq e^{H(p)-H(p_1)+O(n^{(2k-1)\alpha-1})}+e^{-\frac{m^\beta}{8}+H(p)}\,.
\end{equation}
Note that the condition $p_i=O(n^{\alpha-1})$ needed for \eqref{eq:Pk-est} holds since $\beta<1$. Now, using the mean value theorem, for each $i\in\{1,2\}$ there is a $p_i^*$ between $p$ and $p_i$ such that
\[
|H(p)-H(p_i)|=|p_i-p|\cdot |H'(p_i^*)|=O\left(\frac{m^{\frac{(1+\beta)}{2}}}{N}\right)O\left(n^{(k-1)\alpha+1}\right)=O\left(n^{\frac{(1+\alpha)(1+\beta)}{2}+(k-1)\alpha-1}\right)\,.
\]
Now, for $\beta<(k+1)\alpha/(1+\alpha)$, the right hand side in the above will be $O(n^{(3k\alpha-1)/2})$. On the other hand, using $H(p)=O(n^{k\alpha})$, when $\beta>k\alpha/(1+\alpha)$ the term $e^{-m^\beta/8 + H(p)}$ will be $o(1)$. Combining these with Eq. \eqref{eq:sandwich-1}, and choosing $\beta$ in the interval $\left(\frac{k\alpha}{1+\alpha},\frac{(k+1)\alpha}{1+\alpha}\right)$ we have
\[
\frac{\P_{n,m}(A_k)}{\exp\left[-H(m/N)\right]}=\exp\left\{O(n^{\frac{3k\alpha-1}{2}})+O(n^{(2k-1)\alpha-1})\right\}=\exp\left\{O(n^{\frac{3k\alpha-1}{2}})\right\}\,.
\]
Note that, since $\alpha<1/(2k-1)$ then such $\beta$ would be in $(0,1)$ which is needed by Lemma \ref{lem:hoeffding-cor}.
Therefore,
\begin{align*}
\P_{\unif}(G)=\frac{1}{|\gens_{n,m,k}|}&=\frac{1}{{N\choose m}\P_{n,m}(A_k)}=\frac{1}{{N\choose m}\exp\le\{O\le(n^{\frac{3k\alpha-1}{2}}\ri)-\sum_{r=3}^k|\C_r|\le(\frac{m}{N}\ri)^r\ri\}}\,.
\end{align*}
which finishes proof of Lemma \ref{lem:numGraphs} $\square$
\endproof

%
%
\subsection{Proof of Lemma \ref{lem:S1-S2-S3-lbd}}\label{sec:pf-lem-Si-lbd}

Before going into the details we will provide a high level overview of the proof, focusing on $S_1(G)$.

\subsubsection{A High-level Overview of the Proof.}
By definition
\[
S_1(G)=-\sum_{t=0}^{m-1}\E_\pi E_k(G_t^\pi,\pi(t+1))=-\sum_{t=0}^{m-1} \sum_{r=3}^k\sum_{\ell=0}^{r-2}\E_\pi  \left[N_{r,\ell}^{G_{t,ij}}q_t^{r-1-\ell}\right]\,.
\]
The first approximation we use is to change the randomness given by $\pi$.  Since the partial graph $G_t^\pi$ is a uniformly random subgraph of $G$ that has exactly $t$ edges, we can look at $G_\theta$ which is a random subgraph of $G$ that has each edge of $G$ independently with probability $\theta=t/m$.  The subgraph $G_\theta$ has $t$ edges in expectation which makes it a good approximation for $G^\pi_t$. We use this to show that $-\sum_{t=0}^{m-1}\E_\pi E_k(G_t^\pi,\pi(t+1))$ is approximately equal to $-m\,\E_\theta \int_0^1 E_k(G_\theta,(ij))\,d\theta$ where $(ij)$ is a uniformly random edge of $G$. This step is carried out algebraically via Lemma \ref{lem:t/m-rplace-theta}.
Next, note that $E_k(G_t,ij)$ would be approximately sum of the terms $q_t^{r-\ell-1}$ for all pairs $(\gamma,ij)$
where $\gamma$ is in $\C_{r,\ell}(G)$, and $(ij)$ is an edge in $(G\backslash G_\theta)\cap \gamma$. For any fixed $r$, $\ell$ we will see that the sum of all $q_t^{r-\ell-1}$ terms corresponding to such $(\gamma,ij)$ pair is dominated by the cases where $|\gamma\cap G_\theta|=|\gamma\cap G|-1=\ell$; in other words when $(ij)$ is the only edge of $G\cap \gamma$ that is not in $G_\theta$.  This means each cycle $\gamma\in\C_{r,\ell+1}(G)$ would have a (fixed) contribution of $q_t^{r-\ell-1}$ which is why a term $|\C_{r,\ell+1}|$ appears on the right hand side for $S_1$ in Lemma \ref{lem:S1-S2-S3-lbd}(a) (in fact it is $|\C_{r,\ell}|$ for a shifted range $1\leq \ell\leq r-1$).

\subsubsection{Additional Definitions and Lemmas.}

Next, we will state three axillary lemmas that will be used for the proof. But first we introduce an important subset of $\gens_{n,m,k}$. For any graph $G$, denote its maximum degree by $\Delta(G)$. Also, note that $\C_{r,r}(G)$ counts the number of simple cycles of length $r$ that are contained in $G$.
Define the set of graphs $\subG_{n,m,k}$ by,
\[
\subG_{n,m,k}\equiv\gens_{n,m,k}
\cap\left\{ G~\bigg|~\Delta(G)\leq n^{(k+3)\alpha} \right\}
\cap \left(\cap_{s=k+1}^{2k-2}\left\{ G~\bigg|~|\C_{s,s}(G)|\leq n^{(2k-1)(k+1)\alpha} \right\}\right)
\]
The next lemma will show that $\subG_{n,m,k}$ contains almost all of $\gens_{n,m,k}$ and its proof is given in Appendix \ref{app:auxillary}
\begin{lemma}\label{lem:H-isAll}
If $m,n,k$ satisfy conditions of Lemma \ref{lem:S1-S2-S3-lbd} then ${|\subG_{n,m,k}|}\geq \le[1-O(e^{-n^{k\alpha}})\ri]{|\gens_{n,m,k}|}$.
\end{lemma}
We also need to state the following useful upper bound, proved in Appendix \ref{app:auxillary}, on the terms $N_{r,\ell}^{G_t,ij}$ appearing in $S_i$'s.
\begin{lemma}\label{lem:Si-bound}
If $m,n,k$ satisfy conditions of Lemma \ref{lem:S1-S2-S3-lbd} then for all $3\leq r\leq k$ and $G\in\subG_{n,m,k}$ we have
\begin{itemize}
\item[(a)] If $0\leq \ell < r-1$ then $N_{r,\ell}^{G_t^\pi,ij} = O\left(\Delta(G)^\ell n^{r-2-\ell}\right) =O\left(n^{r-2-\ell+\ell(k+3)\alpha}\right)$.
\item[(b)] If $0\leq s < r$ then $|\C_{r,s}(G)|= O\left(\Delta(G)^{s-1}n^{r-s+\alpha}\right)=O\left(n^{r-s+s(k+3)\alpha}\right)$.


\end{itemize}
%
\end{lemma}
%
Before stating the last auxiliary lemma we need to define the following.
\begin{definition} Let $e_1,\ldots,e_s$ be a set of $s$ edges of $G$. Define
$A_{e_1,\ldots,e_s}^{t,\pi}$ to be the event that for all $1\leq
i\leq s:~~e_i\in G_t^\pi$. Similarly, define
$B_{e_1,\ldots,e_s}^{t,\pi}$ to be the event that for all $1\leq
i\leq s:~~e_i\notin G_t^\pi$. Let also $C_{e_i}^{t,\pi}$ be the
event that $\pi(t+1)=e_i$. Also, as a convention (when the index $s=0$ is used) the two sets $A_{\emptyset}^{t,\pi}$
$B_{\emptyset}^{t,\pi}$ contain everything hand have probability $1$.
\end{definition}
\begin{lemma}\label{lem:t/m-rplace-theta}
If $m,n,k$ satisfy conditions of Lemma \ref{lem:S1-S2-S3-lbd} then
for any three integers $a,b,c$ in $\{0,1,\ldots,k\}$ and any set of edges $e_1,e_2,\ldots,e_{a+b+1}$ of $G$ the following hold
\begin{itemize}
\item[(a)]
$
\sum_{t=0}^{m-1}\P_{\pi}\le(A_{e_1,\ldots,e_a}^{t,\pi}\cap
B_{e_{a+1},\ldots,e_{a+b}}^{t,\pi}\ri)(1-\frac{t}{m})^c\leq O(1)+m\int_0^1\theta^a(1-\theta)^{b+c}d\theta
$.

\item[(b)]
$
\sum_{t=0}^{m-1}\P_{\pi}\le(A_{e_1,\ldots,e_a}^{t,\pi}\cap
B_{e_{a+1},\ldots,e_{a+b}}^{t,\pi}\cap
C_{e_{a+b+1}}^{t,\pi}\ri)(1-\frac{t}{m})^c \leq O(\frac{1}{m})+\int_0^1\theta^a(1-\theta)^{b+c}d\theta$.

\item[(c)] $\sum_{t=0}^{m-1}\P_{\pi}\le(A_{e_1,\ldots,e_a}^{t,\pi}\cap B_{e_{a+1},\ldots,e_{a+b}}^{t,\pi}\ri)(1-\frac{t}{m})^c \geq -O(\sqrt{m})+m\int_0^1\theta^a(1-\theta)^{b+c}d\theta$.
\end{itemize}
\end{lemma}
Proof of Lemma \ref{lem:t/m-rplace-theta} is provided in Appendix \ref{app:auxillary}. Next, we prove Lemma \ref{lem:S1-S2-S3-lbd}.

\subsubsection{Finalizing Proof of Lemma \ref{lem:S1-S2-S3-lbd}.}

\proof{Proof of Lemma \ref{lem:S1-S2-S3-lbd} (a).}
Recall that
$S_1(G)=-\sum_{t=0}^{m-1}\sum_{r=3}^k\sum_{\ell=0}^{r-2}\E_{\pi}~N_{r,\ell}^{G_t^\pi,\pi(t+1)}q_t^{r-1-\ell}$,
where $N_{r,\ell}^{G_t^\pi,\pi(t+1)}$ is number of cycles of length $r$
in $K_n$ that include edge $\pi(t+1)$ and have exactly
$\ell$ edges belonging to $G_t^\pi$. Every such cycle, will contain at least $\ell+1$ edges of $G$ so it belongs to $\C_{r,s}(G)$
for some $s$ with $r-1\geq s\geq \ell +1$. This suggests another way to calculate $S_1(G)$. For every cycle that belongs
to $\C_{r,s}(G)$ we can calculate its contribution in $S_1(G)$. Precisely, fix a cycle $\gamma_{r,s}\in \C_{r,s}(G)$.
Let $s_1(\gamma_{r,s})$ be sum of all terms in $S_1(G)$ that are contributed by this cycle. Let $\{e_1,\ldots,e_s\}$ be the set of all $s$ edges in $\gamma_{r,s}\cap G$.
In order for $\gamma_{r,s}$ to be considered in $N_{r,\ell}^{G_t^\pi,\pi(t+1)}$ we need to have $\ell+1$ distinct indices $i_{1},\ldots,i_{\ell+1}$ in $[s]$ such that $\{e_{i_1},\ldots,e_{i_\ell}\}\in G_t^\pi$,  $e_{i_{\ell+1}}=\pi(t+1)$ and $\{e_1,\ldots,e_s\}\backslash \{e_{i_1},\ldots,e_{i_{\ell+1}}\}\in G\backslash (G_t^\pi\cup\{e_{\ell+1}\})$. There are ${s\choose\ell}$ ways to pick the first $\ell$ indices and $(s-\ell)$ ways to pick $e_{\ell+1}$ from the remaining ones. Therefore,
\begin{eqnarray}
s_1(\gamma_{r,s})&=&-\sum_{\ell=0}^{s-1}{s\choose\ell}(s-\ell)\sum_{t=0}^{m-1}\P(A_{e_{i_1},\ldots,e_{i_\ell}}^{t,\pi}\cap C_{e_{i_{\ell+1}}}^{t,\pi}\cap B_{\{e_1,\ldots,e_s\}\backslash \{e_{i_1},\ldots,e_{i_{\ell+1}}\}}^{t,\pi})q_t^{r-1-\ell}\,.\label{eq:s1-eq-1}
\end{eqnarray}
Now, using $q_t=\frac{m-t}{N-t}=(\frac{N}{N-t})(\frac{m}{N})(\frac{m-t}{m})\leq (\frac{N}{N-m})(\frac{m}{N})(\frac{m-t}{m})$,
Eq. \eqref{eq:s1-eq-1}, Lemma \ref{lem:t/m-rplace-theta}(b) for $a=\ell, b=s-(\ell+1), c=r-\ell-1$, and that $N/(N-m)\geq 1$, we have
\begin{eqnarray}
s_1(\gamma_{r,s})&\geq&-\le(\frac{N}{N-m}\ri)^{r-1}~\sum_{\ell=0}^{s-1}{s\choose\ell}(s-\ell)\le(\frac{m}{N}\ri)^{r-1-\ell}\le[O(\frac{1}{m})+\int_0^1\theta^{\ell}(1-\theta)^{r+s-2\ell-2}d\theta\ri]\no\,.
%
\end{eqnarray}
It is easy to see that the summation is dominated by the term $\ell=s-1$ since other terms are an extra factor $m/N$ smaller. The same way, all of the terms involving $O(1/m)$ are smaller by a factor $m$.
Therefore, using $[1+m/(N-m)]^{r-1}=1+O(m/N)$, the largest order term is equal to $-(m/N)^{r-s}s\int_0^1\theta^{s-1}(1-\theta)^{r-s}d\theta$ and everything else is dominated by a constant times $(m/N)^{r-s+1}$; i.e.,
\begin{eqnarray}
s_1(\gamma_{r,s})&\geq&-\le[1+O(\frac{m}{N})\ri]\le(\frac{m}{N}\ri)^{r-s}s\int_0^1\theta^{s-1}(1-\theta)^{r-s}d\theta\no\,.
\end{eqnarray}
Now, considering all possible cycles $\gamma_{r,s}$ we obtain
\[
S_1(G)\geq -\le[1+O(\frac{m}{N})\ri]\sum_{r=3}^k\sum_{s=1}^{r-1}|\C_{r,s}(G)|\le(\frac{m}{N}\ri)^{r-s}s\int_0^1\theta^{s-1}(1-\theta)^{r-s}d\theta\,.
\]
The last step involves simplifying the terms that involve an extra $O(m/N)$ term. In particular,
using Lemma \ref{lem:Si-bound}(b) we have
\begin{align*}
O(\frac{m}{N})\sum_{r=3}^k\sum_{s=1}^{r-1}|\C_{r,s}(G)|\le(\frac{m}{N}\ri)^{r-s}s\int_0^1\theta^{s-1}(1-\theta)^{r-s}d\theta
&= O\left(\sum_{r=3}^k\sum_{s=1}^{r-1}n^{(r-s)+s(k+3)\alpha+(r-s+1)\alpha-(r-s+1)}\right)\\
&= O\left(n^{\alpha(k+3)(k-1)-1}\right)\,.
\end{align*}
This finishes proof of part (a).

\proof{Proof of Lemma \ref{lem:S1-S2-S3-lbd} (b).}
First we need to approximate the number of forbidden pairs $|F(G_t^\pi)|$.
\begin{align}
|F(G_t^\pi)|&=\sum_{(ij)}\ind(\sum_{r=3}^k N_{r,r-1}^{G_t^\pi,ij}>0)\label{eq:F(G)Lower1}\\
&\geq\sum_{r=3}^k\sum_{\gamma\in\C_{r,r-1}(G)} \ind\Big(\gamma\in\C_{r,r-1}(G_t^{\pi})\Big)-\sum_{(ij)}\left[\sum_{r=3}^k N_{r,r-1}^{G_t^\pi,ij} \right]\ind\Big(\sum_{r=3}^k N_{r,r-1}^{G_t^\pi,ij}>1\Big)\no\,,
\end{align}
where the inequality is based on a version of inclusion-exclusion formula. In particular, each edge $(ij)$ with $\sum_{r=3}^k N_{r,r-1}^{G_t^\pi,ij}=1$ is counted exactly once in both sides of the inequality. But the edges $(ij)$ with $\sum_{r=3}^k N_{r,r-1}^{G_t^\pi,ij}>1$ could be counted at most $\sum_{r=3}^k N_{r,r-1}^{G_t^\pi,ij}$ times in the first summation of the right hand side. Next, we are going to show that the second term on the right hand side can be ignored. In particular, the second term is less than the number of times two vertices $i$ and $j$ are connected by two paths of length at most $k-1$ in $G_t^\pi$. This means $i$ and $j$ are two vertices of a cycle of length between $k+1$ to $2k-2$ in $G_t^\pi$ (note that by design $G_t^\pi$ has no cycle of length up to $k$). Since the number of vertices in such cycles is still a constant, we have
\begin{align}
\sum_{(ij)}\left[\sum_{r=3}^k N_{r,r-1}^{G_t^\pi,ij} \right]\ind\Big(\sum_{r=3}^k N_{r,r-1}^{G_t^\pi,ij}>1\Big) = O\left(
\sum_{s=k+1}^{2k-2} |\C_{s,s}(G)|\right)
= O(n^{(2k-1)(k+1)\alpha})\label{eq:F(G)LowerTail}\,,
\end{align}
where the last equality uses $G\in\subG_{n,m,k}$.

On the other hand, for any cycle $\gamma\in \C_{r,r-1}(G)$, using Lemma \ref{lem:t/m-rplace-theta}(c) for $a=r-1, b=c=0$, we have
\[
\sum_{t=0}^{m-1}\E_{\pi}\ind\Big(\gamma\in\C_{r,r-1}(G_t^{\pi})\Big)\geq - O(\sqrt{m})+m\int_{\theta=0}^1\theta^{r-1}d\theta\,.\]
Thus,
\begin{eqnarray}
S_2(G)
&=&\frac{1}{N}\sum_{t=0}^{m-1}\E_{\pi}F(G_t^{\pi})\no\\
&\geq&-O\Big(n^{(2k-1)(k+1)\alpha+\alpha-1}\Big)-\frac{\sqrt{m}}{N}\sum_{r=3}^k|\C_{r,r-1}(G)|+\frac{m}{N}\sum_{r=3}^k|\C_{r,r-1}(G)|\int_{\theta=0}^1\theta^{r-1}d\theta\no\\
&\geq&-O\Big(n^{2k(k+2)\alpha-1}\Big)
-O(n^{\frac{\alpha}{2}-\frac{1}{2}})O(n^{1+(k-1)(k+3)\alpha})
+\frac{m}{N}\sum_{r=3}^k|\C_{r,r-1}(G)|\int_{\theta=0}^1\theta^{r-1}d\theta\label{eq:999}\\
&\geq&-O\Big(n^{k(k+3)\alpha-1/2}\Big)+\frac{m}{N}\sum_{r=3}^k|\C_{r,r-1}(G)|\int_{\theta=0}^1\theta^{r-1}d\theta\,.\no
\end{eqnarray}
%
Here Eq. \eqref{eq:999} uses Lemma \ref{lem:Si-bound}(b). This concludes proof of part (b).

\proof{Proof of Lemma \ref{lem:S1-S2-S3-lbd} (c).}
Recall the set $Q(G_t)$ from \S \ref{sec:alg-main-res}. First note that by definition of $Z(G_t^\pi)$ and $Z_0(G_t^\pi)$ we obtain
\begin{eqnarray}
S_3(G)&=&-\sum_{t=0}^{m-1}\E_\pi\log\le(\frac{\sum_{(ij)\in Q(G_t^{\pi})}\exp\le(-\sum_{r=3}^k\sum_{\ell=0}^{r-2}N_{r,\ell}^{G_t^{\pi},ij}q_t^{r-1-\ell}\ri)}{\sum_{(ij)\in Q(G_t^{\pi})}1}\ri)\,.\no
\end{eqnarray}
Now using $e^{-x}\leq 1-x+\frac{x^2}{2}$ for $x>0$ we have
\begin{eqnarray}
S_3(G)&\ge&-\sum_{t=0}^{m-1}\E_\pi\log\le(1-\sum_{(ij)\in Q(G_t^{\pi})}\frac{\le(\sum_{r=3}^k\sum_{\ell=0}^{r-2}N_{r,\ell}^{G_t^{\pi},ij}q_t^{r-1-\ell}\ri)}{|Q(G_t^{\pi})|}
-\frac{\frac{1}{2}\le(\sum_{r=3}^k\sum_{\ell=0}^{r-2}N_{r,\ell}^{G_t^{\pi},ij}q_t^{r-1-\ell}\ri)^2}
{|Q(G_t^{\pi})|}\ri)\nonumber\,.
\end{eqnarray}
Also note that, using Lemma \ref{lem:Si-bound}(a), we have
\begin{align*}
\sum_{(ij)\in Q(G_t^{\pi})}\frac{\le(\sum_{r=3}^k\sum_{\ell=0}^{r-2}N_{r,\ell}^{G_t^{\pi},ij}q_t^{r-1-\ell}\ri)^2}
{|Q(G_t^{\pi})|}&=O\left(\le[\sum_{r=3}^k\sum_{\ell=0}^{r-2}n^{r-\ell-2+\ell(k+3)\alpha} n^{(r-1-\ell)(\alpha-1)}\ri]^2\right)\\
%
&=O\left(n^{2(k+3)(k-1)\alpha-2}\right)\,,
\end{align*}
and, using a similar argument, each term $\sum_{r=3}^k\sum_{\ell=0}^{r-2}N_{r,\ell}^{G_t^{\pi},ij}q_t^{r-1-\ell}$ is of order $n^{(k+3)(k-1)\alpha-1}$. Therefore, this term and its squared are asymptotically very small (in particular, added together, they are less than $1$). This means we can use $-\log(1-x)\geq x$ for $x<1$ and $|Q(G_t^{\pi})| \leq N$ to obtain
\begin{align}
S_3(G)&\geq\E_\pi\left[\frac{1}{N}\sum_{t=0}^{m-1}\sum_{(ij)\in Q(G_t^{\pi})}\sum_{r=3}^k\sum_{\ell=0}^{r-2}N_{r,\ell}^{G_t^{\pi},ij}q_t^{r-1-\ell}
\right]-m\,O\left(n^{2(k+3)(k-1)\alpha-2}\right)\no\\
&\geq\frac{1}{N}\sum_{t=0}^{m-1}\sum_{r=3}^k\sum_{\ell=0}^{r-2}\E_\pi\left[\sum_{(ij)\in Q(G_t^{\pi})}N_{r,\ell}^{G_t^{\pi},ij}q_t^{r-1-\ell}
\right]-O\left(n^{2k(k+3)\alpha-1}\right).\label{eq:S3geq}
\end{align}
Also, in Eq. \eqref{eq:S3geq}, the summation $\sum_{(ij)\in Q(G_t^{\pi})}$ can be broken to two parts; when $(ij)\in Q(G_t^{\pi})\setminus G$ and when $(ij)\in Q(G_t^{\pi})\cap G$. The latter group is small since, using the same bounds as above, those terms satisfy
\[
\frac{1}{N}\sum_{t=0}^{m-1}\sum_{r=3}^k\sum_{\ell=0}^{r-2}\E_\pi\left[\sum_{(ij)\in Q(G_t^{\pi})\cap G}N_{r,\ell}^{G_t^{\pi},ij}q_t^{r-1-\ell}
\right]=O\left(\frac{m^2n^{(k+3)(k-1)\alpha-1}}{N}\right)=O(n^{(k+3)k\alpha-1})
\]
that can be absorbed in the $O\left(n^{2(k+3)k\alpha-1}\right)$ term of Eq. \eqref{eq:S3geq}.

Now, similar to the proof of (a) we will find contribution of a cycle $\gamma_{r,s}\in\C_{r,s}(G)$ that is denoted by $s_3(\gamma_{r,s})$. The only difference is that this
time the edge $(ij)$ should be part of the $(r-s)$ edges $\gamma_{r,s}\backslash\{e_1,\ldots,e_s\}$ that are not in $G$.
Then we use part (c) of Lemma \ref{lem:t/m-rplace-theta} for $a=\ell$, $b=(s-\ell)$, $c=r-\ell-1$, and $q_t\ge (m/N)(1-t/m)$  to obtain,
\begin{align}
s_3(\gamma_{r,s})&=
\frac{1}{N}\sum_{\ell=0}^{s}{s\choose\ell}(r-s)\sum_{t=1}^{m-1}\P(A_{e_{i_1},\ldots,e_{i_\ell}}^{t,\pi}\cap B_{\{e_1,\ldots,e_s\}\backslash \{e_{i_1},\ldots,e_{i_{\ell}}\}}^{t,\pi})q_t^{r-1-\ell}\no\\
&\geq \frac{1}{N}\sum_{\ell=0}^{s}\le(\frac{m}{N}\ri)^{r-1-\ell}{s\choose\ell}(r-s)\left[m\int_0^1\theta^{s}(1-\theta)^{r+s-2\ell-1}d\theta-O(\sqrt{m})\right]\,.
\end{align}
Similar to part (a), the contribution of $\ell=s$ term will dominate and the remaining terms can be absorbed to the $O(\sqrt{m})$ term.
In particular,
\[
s_3(\gamma_{r,s})
\geq O\left((\frac{m}{N})^{r-s} (r-s)\int_0^1\theta^{s}(1-\theta)^{r-s-1}d\theta\right) - O\left((\frac{m}{N})^{r-s}\sqrt{m}\right)\,.
\]
Therefore,
\[
S_3(G)
\geq\sum_{r=3}^k\sum_{s=0}^{r-2}|\C_{r,s}(G)|\le(\frac{m}{N}\ri)^{r-s}(r-s)\int_0^1\theta^{s}(1-\theta)^{r-s-1}d\theta  - O\left(\sum_{r=3}^k\sum_{s=0}^{r-2}|\C_{r,s}(G)|(\frac{m}{N})^{r-s}m^{-\frac{1}{2}}\right)\,.
\]
Now, using Lemma \ref{lem:Si-bound}(b), we have
\[
O\left(\sum_{r=3}^k\sum_{s=0}^{r-2}|\C_{r,s}(G)|(\frac{m}{N})^{r-s}m^{-\frac{1}{2}}\right)=O(n^{-1/2+(k+3)k\alpha})
\]
which finishes the proof  $\square$

%
%
\section{Running Time of $\algname$ and Proof of Theorem \ref{thm:runtime}}\label{sec:run-time}

In this section we will prove that $\algname$ can be implemented in a way that its expected running time would be of order $n^2m$ operations. The idea is to define surrogate quantities for probabilities $p(ij|G_t)$ that are efficiently computable using sparse matrix multiplications (take order $n^2$ operations per each step of the algorithm). The key point is that, by definition, $p(ij|G_t)$ is a weighted sum over simple cycles. It is known that one can count all cycles (not necessarily simple cycles) of a graph via matrix multiplication of the its adjacency matrix. We will use this fact and prove that the contribution of non-simple cycles will be negligible.

During the execution of $\algname$, after adding $t$ edges, let $\bM_t$ and $\bM_t^{(c)}$ be the adjacency matrices of
the partially constructed graph $G_t$ and its complement $G_t^{(c)}$ respectively. In addition, let $\bQ_t$ be the adjacency matrix
of the graph obtained by all edges $(ij)$ such that $G_t\cup (ij)\in \gens_{n,t+1,k}$. We modify $\algname$ so that it selects the $(t+1)^{th}$
edge from all pairs $(ij)$ with probability $p'(ij|G_t)$ that is equal to $(i,j)$ entry
of the symmetric matrix $\bP'_{G_t}$, defined by
\begin{equation}
\bP'_{G_t}\equiv \le[p'(ij|G_t)\ri]\equiv\frac{1}{Z'(G_t)} \bQ_t\odot\widehat{\exp}\left[-\sum_{r=2}^{k-1}\le(\bM_t+\frac{m-t}{{n\choose2}-t}\bM_t^{(c)}\ri)^r~\right].
\label{eq:P(G_t)matrix}
\end{equation}
Here $Z'(G_t)$ is a normalization constant. Symbols $\odot$ and $\widehat{\exp}$ represent the coordinate-wise multiplication and exponentiation of square matrices. More precisely, for $n\times n$ matrices $\mathbf{A}, \mathbf{B}, \mathbf{C}$ the expression $\mathbf{A}=\mathbf{B}\odot \mathbf{C}$ means that for all $i,j\in [n]$ we have $a_{ij}=b_{ij}c_{ij}$, and similarly
$\mathbf{A}=\widehat{\exp}(\mathbf{B})$ means for all $i,j\in [n]$ we have $a_{ij}=e^{b_{ij}}$. Let us call this modification $\algname'$.

The key result of this section is the following Lemma and is proved in Appendix \ref{app:auxillary}.
\begin{lemma}\label{lem:p(ij|G_t)}
For any non-zero probability term $p'(ij|G_t)$,
\[
p'(ij|G_t) \geq \frac{1}{Z(G_t)}e^{-E_k(G_t,ij)-O\Big(n^{k(k+3)\alpha-2}\Big)}\,,
\]
where $Z(G_t)=\sum_{rs\in Q(G_t)}e^{-E_k(G_t,rs)}$ is the normalization term in definition of $p(ij|G_t)$ from \S \ref{sec:alg-main-res}.
\end{lemma}
Using Lemmas \ref{lem:mainlowerbound} and $\ref{lem:p(ij|G_t)}$ we can see that the output distribution of $\algname'$ still satisfies the inequality $\P_{\alg'}(G)\geq e^{-c_1'n^{-1/2+k(k+3)\alpha}}\P_{\unif}(G)$ for all but $O(e^{-n^{k\alpha}})|\gens_{n,m,k}|$ graphs $G$ in $\gens_{n,m,k}$. More formally, a variant of Lemma \ref{lem:mainlowerbound} holds for $\P_{\alg'}$ using Lemma \ref{lem:mainlowerbound} for $\P_{\alg}$ and Lemma \ref{lem:p(ij|G_t)}. Next, we focus on the implementation of $\algname'$.

The fact that $\algname'$ has polynomial running time is clear since the matrix of the probabilities at any step, $\bP_{G_t}$, can be calculated using matrix multiplication. In fact a myopic calculation shows that $\bP_{G_t}$ can be calculated with $O(k\,n^3)=O(n^3)$ operations. This is because $r^{th}$ power of a matrix for any $r$ takes $O(rn^3)$ operations to compute. So we obtain the simple bound of $O(n^3m)$ for the running time. But we can improve this running time by at least a factor $n$ with exploiting the structure of the matrices.

Notice that the adjacency matrix $\bQ_t$ is equal to $\mathbf{J}_n-\widehat{{\rm sign}}(\sum_{r=0}^{k-1}\bM_t^r)$
where $\mathbf{J}_n$ is the $n$ by $n$ matrix of all ones and the $\widehat{{\rm sign}}(\mathbf{B})$ for any matrix $\mathbf{B}$ means the \emph{sign} function is applied to each entry of $B$.
This is correct since any \emph{bad} pair $(ij)$, that cannot be added to $G_t$, corresponds to a path in $G_t$ of length $r$ between $i$ and $j$ for $0\leq r\leq k-1$. Such path forces the $ij$ entry of the matrix $\bM_t^r$ to be positive.

Now we can store the matrices $\bM_t,\ldots,\bM_t^{k-1}$ at the end of each iteration and use them to efficiently calculate $\bM_{t+1},\ldots,\bM_{t+1}^{k-1}$.  This is because the differences $\bM_{t+1}-\bM_t$
are sparse matrices and updating the matrix multiplications can be done with $O(n^2)$. More precisely, we can use
\begin{eqnarray*}
\bM_{t+1}^{r}&=&\left[\bM_t+(\bM_{t+1}-\bM_t)\right]^r=\bM_t^r + \mathbf{L}\,,
\end{eqnarray*}
where $\mathbf{L}$ is a linear sum of matrix products where each term contains at least one of $(\bM_{t+1}-\bM_t),\cdots,(\bM_{t+1}-\bM_t)^{r-1}$. Since $\bM_{t+1}-\bM_t$ has $O(1)$ non-zero entries then the total operations required for calculating $\mathbf{L}$ is of $O(n^2)$. A similar argument can be used for calculating
$
\Big[\bM_{t+1}+\frac{m-{t+1}}{{n\choose2}-{t+1}}\bM_{t+1}^{(c)}\Big]^r
$
using sparsity of both $\bM_{t+1}-\bM_t$ and $\bM_{t+1}^{(c)}-\bM_t^{(c)}$.

Since Theorem \ref{thm:main} shows that $\algname$ and hence $\algname'$ are successful with probability $1-n^{-1/2+k(k+3)\alpha}$, the expected running-time of $\algname'$ for generating an element of $\gens_{n,m,k}$ is also $O(n^2m)$, for $n$ large enough, which finishes proof of Theorem \ref{thm:runtime} $\square$

%
%

\section{Comparing $\algname$ and $C_k$-free Process}\label{sec:compare-w-TF}

In this section, we perform a theoretical (\S \ref{subsec:compare-w-cl-free}) and an empirical comparison (\S \ref{subsec:empirical-compare-w-cl-free}) between our results for $\algname$ and existing theory for $C_k$-free process.
The motivation for this comparison is due to recent research by \cite{Morris2013,BohmanKeevash2013}. They show that certain graph parameters in the $C_3$-free process concentrate around their value in uniformly random $C_3$-free graphs. But these papers do not provide any formal statement on closeness of the two distributions. Our goal is to understand how close the output distribution of $C_3$-free and $\algname$ are to the uniform distribution on $\gens_{n,m,k}$.

\subsection{Concentration Inequality for Graph Parameters}\label{subsec:compare-w-cl-free}

Recall that $Q(G)$ was defined to be the subset of edges in $K_n$ that adding them to $G$ does not create a cycle of length at most $k$. We enrich this notation by adding a subscript $k$, i.e. using $Q_k(G)$. Also let $\tf$ be the short notation for the triangle-free ($C_3$-free) process. We will show that Theorem \ref{thm:main} provides a sharper concentration than Theorem 2.1 of \cite{Morris2013} for $Q_3(G)$. \cite{Morris2013} show that
\begin{align}
\lim_{n\to\infty}\P_{\tf}\left\{~\left|1-\frac{|Q_{3}(G)|}{\E_\unif|Q_{3}(G)|}\right| < 2e^{2m^2/n^3}n^{-1/4}(\log n)^3\right\}=1\,.\label{eq:Q(G)BoundByThem}
\end{align}
On the other hand, we note the following corollary of Theorem \ref{thm:main} for $Q_k(G)$ that is proved in  Appendix \ref{app:auxillary}.
\begin{corollary}\label{cor:Q(G)BoundByUs}
Let $n$, $m$, and $k$ satisfy the conditions of Theorem \ref{thm:main}. Then there exists a constant $c_3$ such that
\begin{align}
\P_{\alg}\left\{~~\left|1-\frac{|Q_{k}(G)|}{\E_{\unif}|Q_{k}(G)|}\right| < c_3 n^{-1+(2k-1)(k+1)\alpha}\right\} = 1-O(n^{-1/2+k(k+3)\alpha}) \,.\label{eq:Q(G)BoundByUS}
\end{align}
\end{corollary}
For small enough $\alpha$, the bound \eqref{eq:Q(G)BoundByUS} is clearly more general than \eqref{eq:Q(G)BoundByThem} since it applies to $k\ge3$ and the rate of convergence for the probability is provided. But, more importantly, the error term $n^{-1+(2k-1)(k+1)\alpha}$ is much smaller than $2e^{2m^2/n^3}n^{-1/4}(\log n)^3\approx n^{-1/4}$ when $(2k-1)(k+1)\alpha<3/4$. For example, when $k=3$ and $\alpha<0.025$, the error term in \eqref{eq:Q(G)BoundByUS} is $O(n^{-1/2})$. We should note that the result of \cite{Morris2013} is instead valid for a much larger range of graphs (up to $m\approx n^{1.5}$) compared to our bound that is valid for $m=O(n^{1+\alpha})$.

\cite{Morris2013} also prove similar asymptotic approximations as in \eqref{eq:Q(G)BoundByThem} for several other graph parameters than $|Q(G)|$. We expect the same argument as above can be applied to obtain sharper concentrations for those parameters as well (when $\alpha$ is a small).

It is worth noting that the above comparison is between the bounds proved for two different algorithms, $C_k$-free process and $\algname$. But an interesting comparison, that we leave for future research, could be done by applying the analysis of $\algname$ from this paper to $C_k$-free process and obtaining a similar variant of \eqref{eq:Q(G)BoundByUS} for the $C_k$-free process.

%
%

\subsection{Empirical Comparison}\label{subsec:empirical-compare-w-cl-free}

In last section we showed that our bound on $\dv(\P_{\alg},\P_{\unif})$ is sharper than existing theory on closeness of $C_3$-free process to $\P_{\unif}$. But we did not answer the question: Is $\dv(\P_{\alg},\P_{\unif})$ is smaller than $\dv(\P_{\tf},\P_{\unif})$. In order to shed light on this, below we perform an empirical comparison between $\algname$ and triangle-free process.

Given that at step $t$ of either algorithm we know the value of $p(\pi(t+1)|G^{\pi}_t)$, we can use that to (empirically) compare the output distribution of each algorithm with uniform. In particular, for a successful run of $\algname$ that outputs a graph $G$ with ordering $\pi$ of its edges we estimate its \emph{multiplicative bias} by
\begin{align}
 \textrm{Bias}^{\pi}_{\alg}\equiv \frac{m!\,\prod_{t=0}^{m-1}p(\pi(t+1)|G^{\pi}_t)}{\left\{{N\choose m}\exp\le[-{n\choose 3}\le(\frac{m}{N}\ri)^3\ri]\right\}^{-1}}\,. \label{eq:def_bias}
\end{align}
From Lemma \ref{lem:numGraphs}, for $\alpha<\min[1/(2k-1),1/(3k)]\approx 0.11$, the denominator in $ \textrm{Bias}^{\pi}_{\alg}$ is close to $\P_{\unif}(G)$ and the numerator is approximately equal to $\P_{\alg}(G)$ since there are $m!$ orderings $\pi$ for edges of $G$. Similarly, we can define  $\textrm{Bias}^{\pi}_{\tf}$ by using the values $p(\pi(t+1)|G^{\pi}_t)$ from the triangle-free process. Therefore, $\textrm{Bias}^{\pi}_{\alg}$ and $ \textrm{Bias}^{\pi}_{\tf}$ are approximations to $\P_{\alg}/\P_{\unif}$ and $\P_{\tf}/\P_{\unif}$ respectively. In other words, if the multiplicative bias of an algorithm is closer to $1$ then its output distribution is also closer to uniform.

\begin{figure}[t]
\centering
\begin{tabular}{cc}
\includegraphics[width=0.45\linewidth]{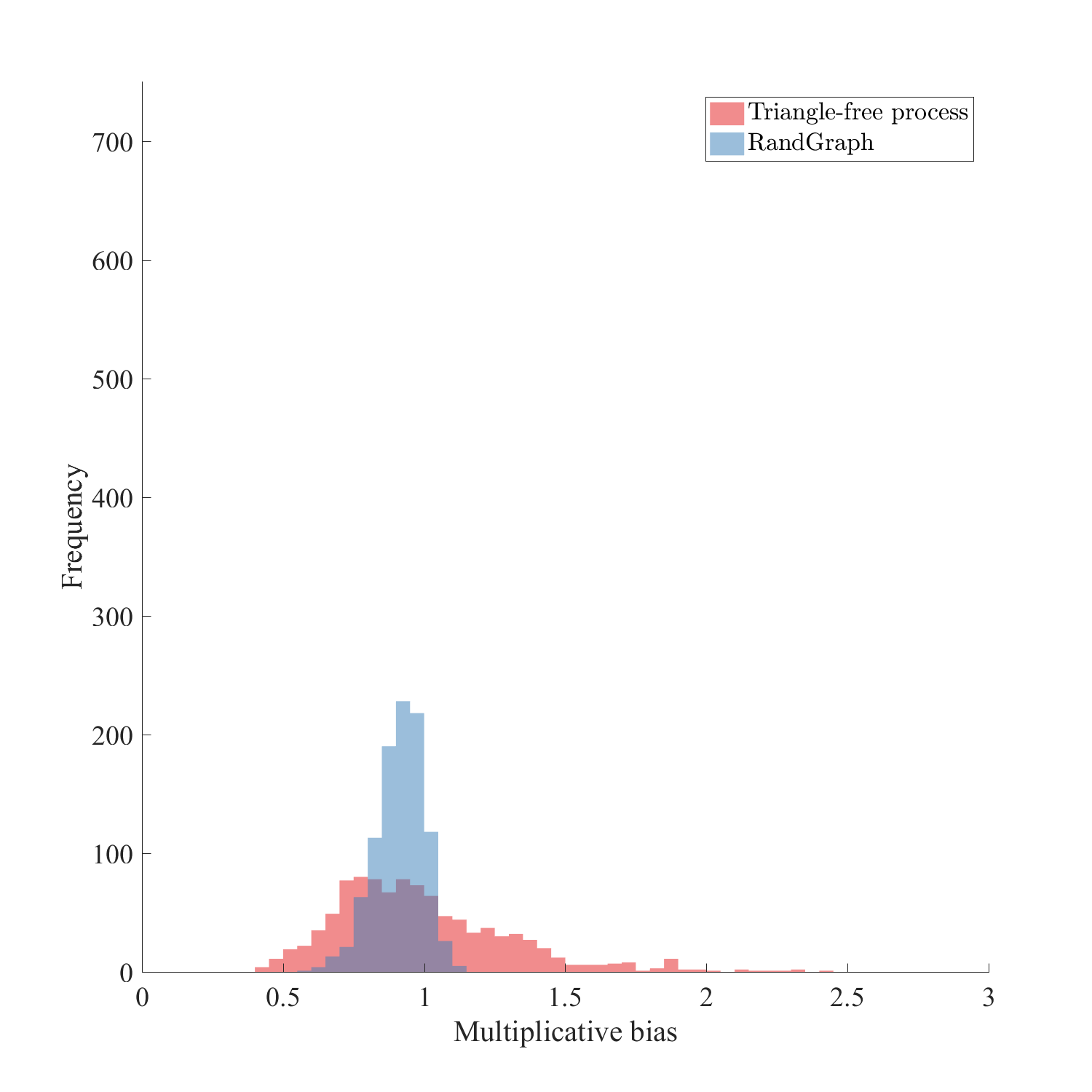} &
\includegraphics[width=0.45\linewidth]{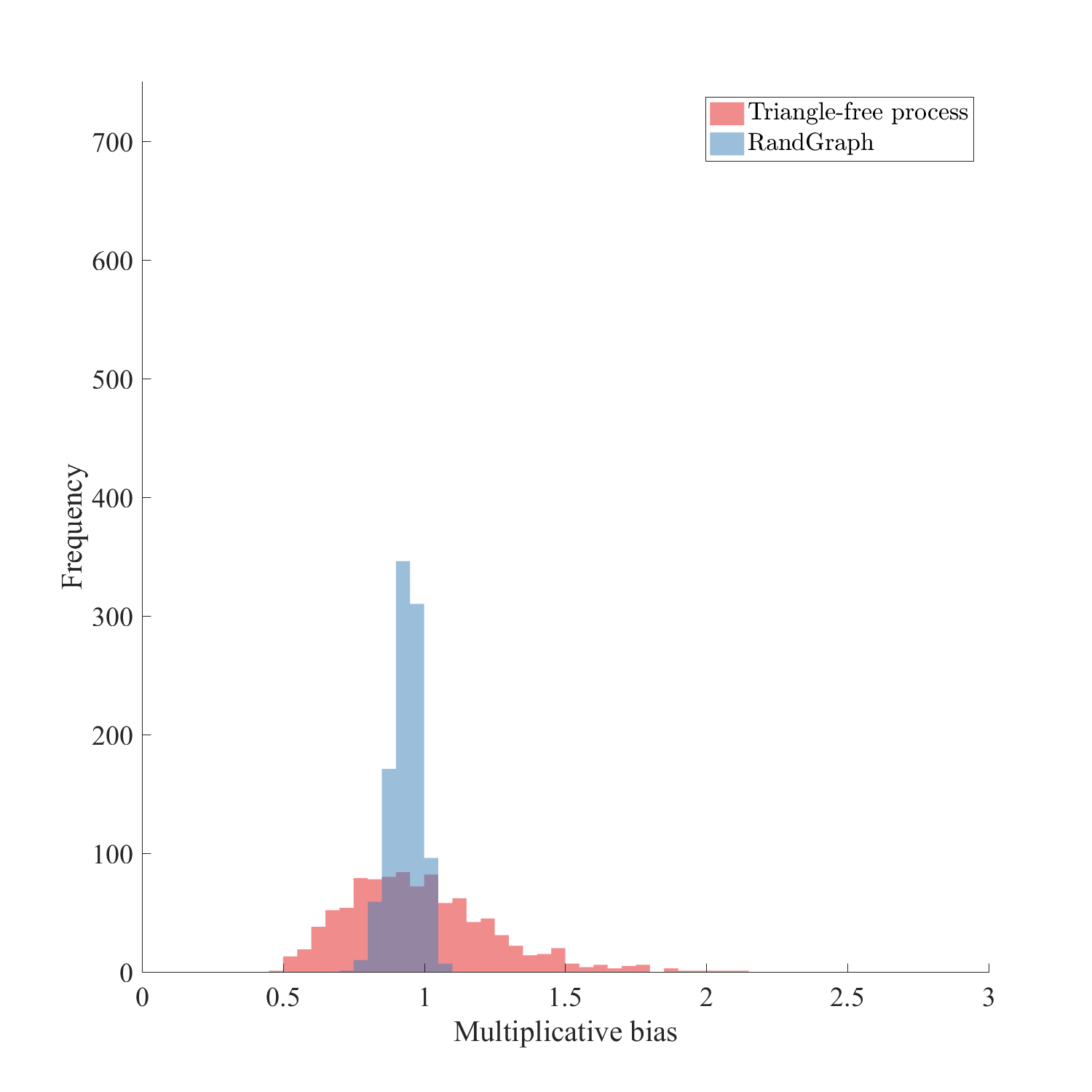} \\
(a) $n=50$ & (b) $n=100$\\
&\\
\includegraphics[width=0.45\linewidth]{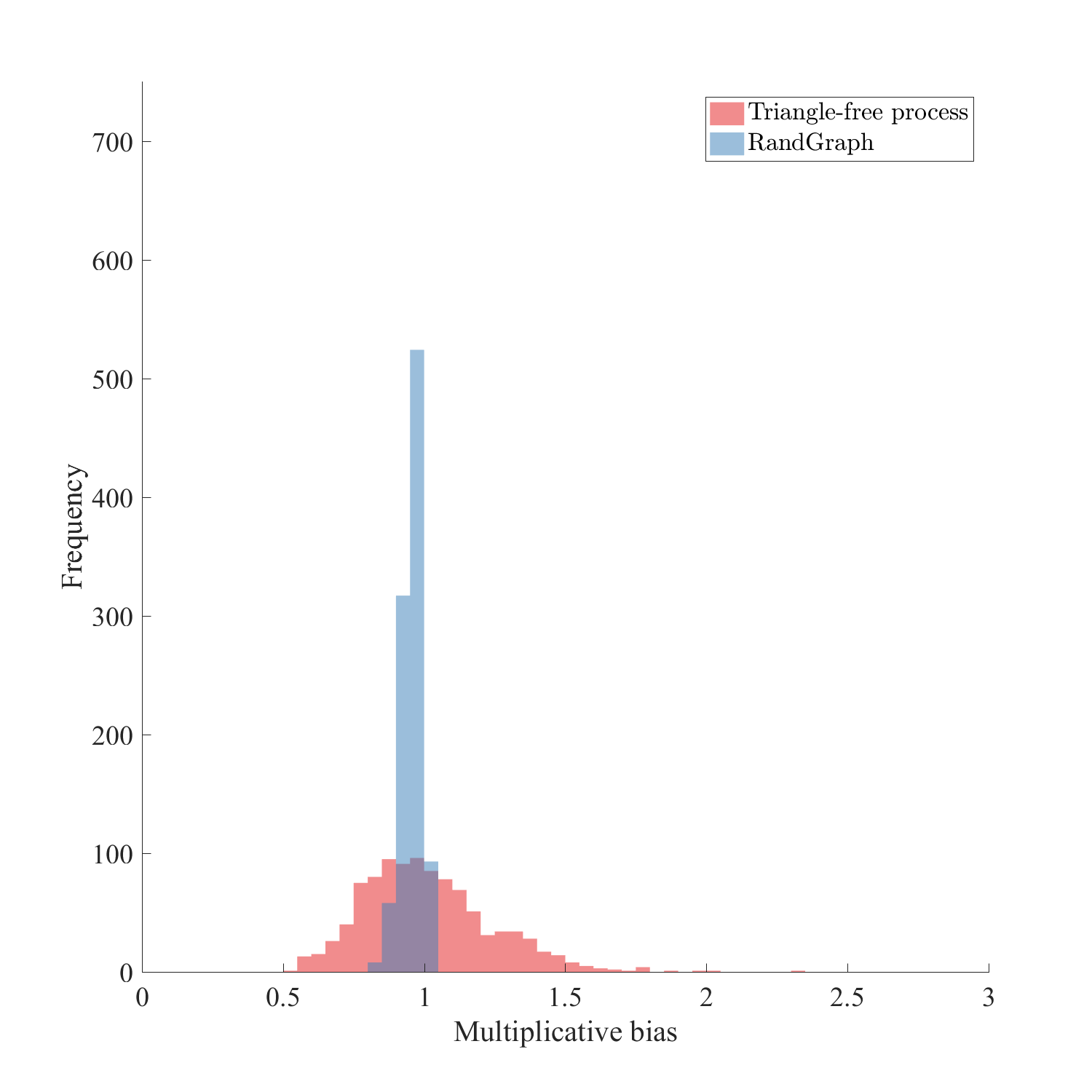} &
\includegraphics[width=0.45\linewidth]{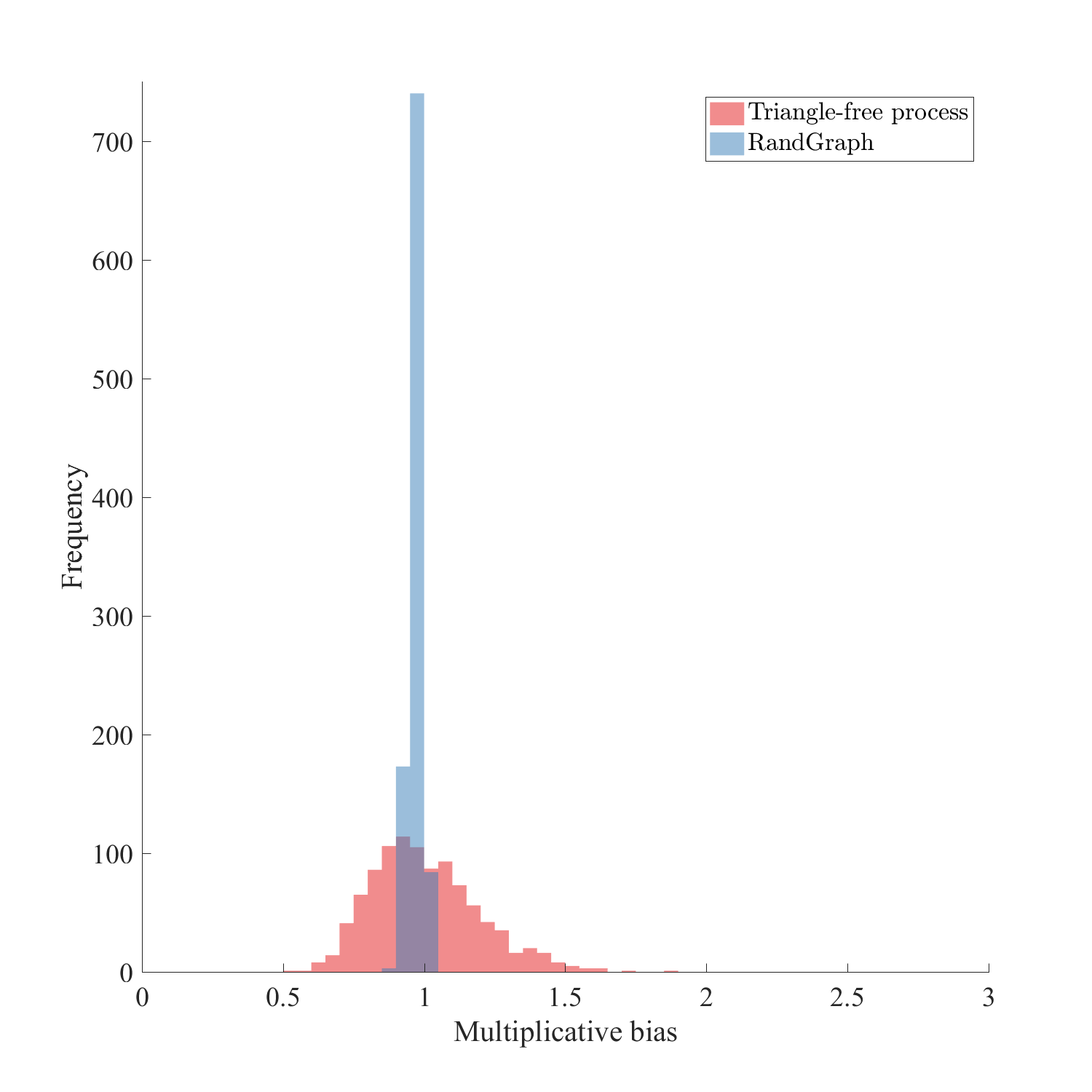} \\
(a) $n=200$ & (b) $n=400$
\end{tabular}
\caption{Histogram of multiplicative bias for 1,000 runs of $\algname$ and triangle-free process (i.e.,  $\textrm{Bias}^{\pi}_{\alg}$ and $ \textrm{Bias}^{\pi}_{\tf}$) for $n\in\{50, 100,200,400\}$. In all cases $m=n^{1+\alpha}$ with $\alpha=0.1$.}
\label{fig:empirical_comparison}
\end{figure}

Next, for $n$ in $\{50, 100, 200,400\}$ and $m=n^{1+\alpha}$ where $\alpha= 0.1$, we execute $\algname$ and triangle-free process $1,000$ times. First we note that no algorithm failed during the 1,000 repetitions. Figure \ref{fig:empirical_comparison} shows the histograms of $ \textrm{Bias}^{\pi}_{\alg}$ and  $\textrm{Bias}^{\pi}_{\tf}$ for each $n$. 
The following observations can be made from the simulation:
\begin{itemize}
\item Bias values for $\algname$ are more concentrated around $1$ than the ones by triangle-free process. This supports the fact that the distance between $\P_{\alg}$ and $\P_{\unif}$ is less than the distance between $\P_{\tf}$ and $\P_{\unif}$.

\item The bias of $\algname$ seems to converge to $1$ as $n$ grows which suggests that our results (possibly) hold for a larger range of $\alpha$ than what is required by Theorem \ref{thm:main}, i.e., $\alpha\in (0,0.11)$ versus $\alpha\in(0,0.027)$.

\end{itemize}

%
%

\section{Extension to Bipartite Graphs with Given Degrees}\label{sec:application}

The ideas described in \S \ref{sec:idea} can be used to generate random bipartite
graphs with given node degrees. Such graphs define the standard model
for irregular LDPC codes. In this section we will show how to modify $\algname$ for this application.
The analysis of this extension is somewhat cumbersome and is beyond the scope of this paper but we expect it to be conceptually similar to the analysis of $\algname$. Since this is a short section, the notation introduced here is not presented in Table \ref{tab:notations}.

Consider two ordered sequences of positive integers $\bar{r}=(r_1,\ldots,r_{n_1})$ and $\bar{c}=(c_1,\ldots,c_{n_2})$ for degrees of the vertices such that $m=\sum_{i=1}^{n_1}r_i=\sum_{j=1}^{n_2}c_j$.  We would like to generate a random bipartite graph $G(V_1,V_2)$, $V_1=[n_1]$ and $V_2=[n_2]$, with girth greater than $k$ and with degree sequence $(\bar{r}, \bar{c})$. We also assume that $k$ is an even number. Denote the set of all such graphs by $\gens_{\bar{r},\bar{c},k}$. The algorithm is a natural generalization of $\algname$ where the probabilities $p(ij|G_t)$ are adjusted properly.
\begin{algorithm}
\SingleSpacedXI
\begin{algorithmic}
\State \textbf{Input:} Degree sequence $(\bar{r}, \bar{c})$ and $k$
\State \textbf{Output:} An element of $\gens_{\bar{r},\bar{c},k}$ or {\sf FAIL}
\State set $G_0$ to be a graph over vertex sets $V_1=[n_1]$, $V_2=[n_2]$ and with no edges.
\State let $\hat{r}=(\hat{r}_1,\ldots,\hat{r}_n)$ and $\hat{c}=(\hat{c}_1,\ldots,\hat{c}_m)$ be ordered sets that are initialized by $\hat{r}=\bar{r}$ and $\hat{c}=\bar{c}$
\For {each $t$ in $\{0,\ldots,m-1\}$}
\If {adding any edge to $G_t$ creates a cycle of length at most $k$}
\State stop and return {\sf FAIL}
\Else
\State sample an edge $(ij)$ from $V_1\times V_2$ with probability $p''(ij|G_t)$, defined by Eq. \eqref{eq:q(ij|Gt)}
\State set $G_{t+1}= G_t\cup (ij)$
\State set $\hat{r}_i=\hat{r}_i-1$  and $\hat{c}_j=\hat{c}_j-1$
\EndIf
\EndFor
\If {the algorithm does not {\sf FAIL} before $t=m-1$}
\State return $G_m$
\EndIf
\end{algorithmic}
\caption{$\bipalgname$.}
\end{algorithm}

Here each probability $p''(ij|G_t)$ is an approximation to the probability that a uniformly random extension of graph $G_t\cup(ij)$ has girth larger than $k$ (the intuitive reason for this is described in \S \ref{sec:idea}). The estimation procedure for $p''(ij|G_t)$ is slightly more involved than the one used for $p(ij|G_t)$. It relies on considering a \emph{configuration model} representation for the graphs with degree sequence $(\bar{r}, \bar{c})$, see
\citep{BenderC78,bollobas80} for more details on configuration model. Then, building on the idea discussed in \S \ref{sec:idea}, we get the following Poisson-type approximation for $p''(ij|G_t)$,
\begin{eqnarray}
p''(ij|G_t) \equiv \frac{\hat{r}_i\hat{c}_je^{-E_k''(G_t,ij)}}{Z''(G_t)}\,,\label{eq:q(ij|Gt)}
\end{eqnarray}
where $Z''(G_t)$ is a normalization term, and $\hat{r}_i$ $\hat{c}_j$, denote the remaining degrees of $i$ and $j$.
 Furthermore, $E_k''(G_t,ij)\equiv\sum_{r=1}^{k/2}\sum_{\stackrel{\gamma\in \C_{2r}}{(ij)\in \gamma}} p_{ij}^t(\gamma)
$, where $\C_{2r}$ is the set of all simple cycles of length $2r$ in the complete bipartite graph on vertices of $V_1$ and $V_2$ Also, $p_{ij}^t(\gamma)$ is approximately the probability that $\gamma$ is in a random extension of $G_t$ to a random bipartite graph with degree sequence $(\bar{r},\bar{c})$. More precisely,
\[
p_{ij}^t(\gamma)=\frac{(m-t-2r+|\gamma\cap G_t|)!\prod_{\ell\in\gamma\cap V_1}R_{ij}^t(\ell,\gamma)\prod_{\ell\in\gamma\cap V_2}C_{ij}^t(\ell,\gamma)}{(m-t-1)!}\,,
\]
where
 \[
R_{ij}^t(\ell,\gamma)=\left\{
 \begin{array}{llll}
\hat{r}_\ell(\hat{r}_\ell-1)&&&\textrm{If } \deg_{\ell}\Big(\gamma\cap \big[G_t\cup(ij)\big]\Big)=0\,,\\
\hat{r}_\ell&&&\textrm{If } \deg_{\ell}\Big(\gamma\cap \big[G_t\cup(ij)\big]\Big)=1\,,\\
1&&&\textrm{If } \deg_{\ell}\Big(\gamma\cap \big[G_t\cup(ij)\big]\Big)=2\,.
 \end{array}
 \right.
 \]
 Similarly,
  \[
C_{ij}^t(\ell,\gamma)=\left\{
 \begin{array}{llll}
\hat{c}_\ell(\hat{c}_\ell-1)&&&\textrm{If } \deg_{\ell}\Big(\gamma\cap \big[G_t\cup(ij)\big]\Big)=0\,,\\
\hat{c}_\ell&&&\textrm{If } \deg_{\ell}\Big(\gamma\cap \big[G_t\cup(ij)\big]\Big)=1\,,\\
1&&&\textrm{If } \deg_{\ell}\Big(\gamma\cap \big[G_t\cup(ij)\big]\Big)=2\,.
 \end{array}
 \right.
 \]
Here the notation $\deg_v(H)$ for a node $v$ of graph $G$ and subgraph $H$ of $G$ refers to the induced degree of $v$ in $H$.

%
\begin{APPENDICES}

\section{Proofs of Auxiliary Lemmas}\label{app:auxillary}

\proof{Proof of Lemma \ref{lem:Cr0/Cr}}

It is easy to see that $|\C_r|=\textrm{constant}\cdot n^{r}$. Now we try to find an upper bound for the number of paths of length $r$ that intersect at least one edge of $G$. The number of paths $\gamma$ that intersect a fixed edge $(ij)$ in $G$ is of order $O(n^{r-2})$ since there are ${n-2\choose r-2}$ ways to pick the remaining $r-2$ vertices of $\gamma$ and this is the dominating term. And Therefore,
\begin{align*}
\frac{|\C_r\backslash \C_{r,0}(G)|}{|\C_r|}&=O\left(\frac{\sum_{(ij)\in G}n^{r-2}}{n^r}\right)\\
&=O\left(mn^{-2}\right)=O\left(n^{\alpha-1}\right)\,\square
\end{align*}
\endproof

\proof{Proof of Lemma \ref{lem:monotonicity_inequalities}}

We note that for any $0<p<1$, the random graph model $\gens(n,p)$ is equivalent to the random graph model $\gens_{n,m}$ conditioned on $m(G)=m$. Thus, for a random graph $G$ we have
\begin{align*}
\P_{n,p}(A_k)&=\P_{n,p}\Big(A_k\cap\{\nedge(G)\geq m\}\Big)+\P_{n,p}\Big(A_k\cap\{\nedge(G)<m\}\Big)\\
%
%
&\leq \sum_{\ell=m}^N \P_{n,p}\Big(A_k\big|\nedge(G)=m\Big)\P_{n,p}\Big(\nedge(G)=\ell\Big)
+\P_{n,p}\bigg(\nedge(G)<m\bigg)\\
&\leq \P_{n,p}\Big(A_k\big|\nedge(G)=m\Big)\sum_{\ell=m}^N \P_{n,p}\Big(\nedge(G)=\ell\Big)
+\P_{n,p}\bigg(\nedge(G)<m\bigg)\\
&\leq \P_{n,m}(A_k) + \P_{n,p}\bigg(|\nedge(G)|<m\bigg)\,,
\end{align*}
where the second inequality uses monotonicity of property $A_k$. Similarly,
\begin{align*}
\P_{n,p}(A_k)&\geq\P_{n,p}\Big(A_k\cap\{\nedge(G)\leq m\}\Big)\\
&=\sum_{\ell=0}^m \P_{n,p}\Big(A_k\big|\nedge(G)=\ell\Big) \P_{n,p}\Big(\nedge(G)=\ell\Big)\\
&\geq\P_{n,p}\Big(A_k\big|\nedge(G)=m\Big)\sum_{\ell=0}^m  \P_{n,q}\Big(\nedge(G)=\ell\Big)\,,~~~\textrm{using monotonicity of $A_k$}\\
&=\P_{n,m}(A_k)\P_{n,p}\Big(\nedge(G)\leq m\Big)\\
&=\P_{n,m}(A_k)-\P_{n,p}\Big(\nedge(G)> m\Big)\,.
\end{align*}

\endproof

\proof{Proof of Lemma \ref{lem:hoeffding-cor}}
First we state the following modified version of Hoeffding inequality, adapted from Corollary 3.2 in \citep{StW99}.
\begin{proposition}[Hoeffding inequality]\label{prop:hoeffding}
Let $X_1,\ldots,X_n$ be independent variables with $0\leq X_i \leq 1$ for all $i\in[n]$, and let $X=\sum_{i=1}^n X_i$. Then for $\delta \leq 4/5$,
\[
\P\left[\,\big|X-\E(X)\big|>\delta\, \E(X) \,\right] \leq e^{-\delta^2\E(X)/4}\,.
\]
\end{proposition}
We can now take $N$ iid Bernoulli$(p)$ random variables corresponding to the potential edges of $G$ in $\gens_{n,p}$ and use Proposition \ref{prop:hoeffding} to obtain, for any
$0<p<1$ and $0<\delta<4/5$,
\[
\P_{n,p}\le(\,\big| \nedge(G)-Np\big|>\delta Np\,\ri)\leq e^{-\delta^2Np/4}\,.
\]
Now we can see that by taking $\delta=\frac{m^{(1+\beta)/2}}{m-m^{(1+\beta)/2}}$,
when $\beta\in (0,1)$ and $m$ is large enough, we have
$\delta<4/5$, $(1+\delta)Np_1=m$, and $\delta^2Np_1\geq m^{\beta}/2$ which give
\begin{align*}
\P_{n,p_1}\Big(\,\nedge(G)>m\Big)&\leq \P_{n,p_1}\Big(\,\nedge(G)>(1+\delta)Np_1\,\Big)\leq e^{-\delta^2Np_1/4}\leq e^{-m^\beta/8}\,.
\end{align*}
For the second inequality, $\P_{n,p_2}\Big(\,\nedge(G)<m\Big)\leq e^{-m^\beta/8}$,
we take $\delta=\frac{m^{(1+\beta)/2}}{m+m^{(1+\beta)/2}}$, which gives $(1-\delta)Np_2=m$ and $\delta^2Np_2\geq m^\beta/2$ and the result similarly follows $\square$


\proof{Proof of Lemma \ref{lem:H-isAll}}

First, we will find an upper bound for probability of the event $\Delta(G)>n^{(k+3)\alpha}$ and a separate bound for the event $\sum_{s=k+1}^{2k-2}|\C_{s,s}(G)|> n^{2k\alpha}$. Then we combine them via union bound.

For maximum degree, we use the following version of Chernoff inequality, Theorem A.1.18 in \citep{AlS92}.
For i.i.d. Bernoulli random variables $X_1,\ldots,X_N$ with mean $p$
\[
\P\le(\sum_{i=1}^N X_i>\eta+Np\ri)<e^{-2\eta^2}\,.
\]
Now combining this with a union bound, for graphs $G$ in $\gens_{n,m,k}$ we have for any $p\in(0,1)$
\begin{eqnarray*}
\P_{n,p}\Big[\Delta(G)>(n-1)p+\eta\Big]&<&ne^{-2\eta^2}\,.
\end{eqnarray*}
Note that the
event $\{\Delta(G)>(n-1)\,p + \eta\}$
is a monotone property (see beginning of \S \ref{sssec:Pk-via-Janson} for definition) but in the opposite direction as $A_k$ that is adding edges to $G$ maintains the property. Therefore, similar to the proof of Lemma \ref{lem:numGraphs} we can take $p_2=\frac{m+m^{\frac{1+\beta}{2}}}{N}$ and obtain
\begin{align*}
\P_{n,m}\Big[\Delta(G)>(n-1)p_2+\eta\Big]&<\P_{n,p_2}\Big[\Delta(G)>(n-1)p_2+\eta\Big]+\P_{n,p_2}\Big[\nedge(G)<m\Big]\\
&< ne^{-2\eta^2}+e^{-\frac{m^\beta}{8}}\,.\no
\end{align*}
Thus, for $\beta=1/2$ and $\eta=n^{\frac{(k+2)\alpha}{2}}$, combining the above bounds with $np_2 = O(n^\alpha)$ and $m^\beta/8>2n^{(k+2)\alpha}$ we have
\begin{equation}
\label{eq:Pm(Hc)-bd}
\P_{n,m}\Big[\Delta(G)>n^{(k+3)\alpha}
\Big]<e^{-n^{(k+1)\alpha}}\,.
\end{equation}

Next, we will find a similar bound for $\P_{n,p}[\sum_{s=k+1}^{2k-2}|\C_{s,s}(G)|> n^{2k\alpha}]$. For this, we use the following concentration inequality for $|\C_{s,s}(G)|$ in $\gens_{n,p}$ that is adapted from Corollary 6.2 of \cite{Vu-extension},
\begin{align}
\P_{n,p}\left[|\C_{s,s}(G)|> \E_{n,p}|\C_{s,s}(G)| + n^{s(k+1)\alpha}\right]=O(e^{-n^{(k+1)\alpha}})\,.
\label{eq:cycle-count-via-KimVu}
\end{align}
In fact, Corollary 6.2 of \cite{Vu-extension} provides a bound for more general subgraph counts (not necessarily cycle counts). But in Vu's bound the tail is of order $\E_{n,p}|\C_{s,s}(G)|=O(n^{s\alpha})$ and the probability is of order $\exp(-n^{\alpha})$. However, we require a smaller probability of order $\exp(-n^{(k+1)\alpha})$ and can afford to pick a larger tail. By choosing $\lambda=4an^{\alpha(k+1)}$ instead of $\lambda=an^{\alpha}$, and leaving everything else unchanged in Vu's proof, all conditions satisfy and we obtain \eqref{eq:cycle-count-via-KimVu}. Therefore,
\begin{align*}
\P_{n,p}\Big[\sum_{s=k+1}^{2k-2}|\C_{s,s}(G)|> n^{(2k-1)(k+1)\alpha}\Big]
&\leq
\P_{n,p}\Big[\sum_{s=k+1}^{2k-2}|\C_{s,s}(G)|> \sum_{s=k+1}^{2k-2}\left(\E_{n,p} |\C_{s,s}(G)|+n^{s(k+1)\alpha}\right) \Big]\\
&\leq \sum_{s=k+1}^{2k-2}\P_{n,p}\Big[|\C_{s,s}(G)|> \E_{n,p} |\C_{s,s}(G)|+n^{s(k+1)\alpha}\Big]\\
&= O(e^{-n^{(k+1)\alpha}})\,.
\end{align*}
Now, defining $p_2$, $m$, and $\beta$ the same as above and repeating the same argument for the monotone property $\sum_{s=k+1}^{2k-2}|\C_{s,s}(G)|> n^{(2k-1)(k+1)\alpha}$ we have
\begin{align*}
\P_{n,m}\Big[\sum_{s=k+1}^{2k-2}|\C_{s,s}(G)|> n^{(2k-1)(k+1)\alpha}\Big]&<\P_{n,p_2}\Big[\sum_{s=k+1}^{2k-2}|\C_{s,s}(G)|> n^{(2k-1)(k+1)\alpha}\Big]+\P_{n,p_2}\Big[\nedge(G)<m\Big]\\
&= O(e^{-n^{(k+1)\alpha}})\,.\no
\end{align*}

Finally, note that in \S \ref{ssec:est-Gnmk} we explicitly calculated $\P_{n,m}(A_k)$ which shows that $\P_{n,m}(A_k)^{-1}$ is of order $e^{O(n^{k\alpha})}$. Hence,
\begin{eqnarray*}
\frac{|\subG_{n,m,k}|}{|\gens_{n,m,k}|}&=&\P_{n,m}\left(\Big[\Delta(G)\leq n^{(k+3)\alpha}\Big]\cap\Big[\sum_{s=k+1}^{2k-2}|\C_{s,s}(G)|\leq n^{(2k-1)(k+1)\alpha}\Big]\,\bigg|\,G\in\gens_{n,m,k}\right)\no\\
&&\\
&=&\frac{\P_{n,m}\left(\Big[\Delta(G)\leq n^{(k+3)\alpha}\Big]\cap\Big[\sum_{s=k+1}^{2k-2}|\C_{s,s}(G)|\leq n^{(2k-1)(k+1)\alpha}\Big]\cap A_k\right)}{\P_{n,m}(A_k)}\no\\
&&\\
&\geq&\frac{\P_{n,m}(A_k)-\P_{n,m}\Big[\Delta(G)>n^{(k+3)\alpha}\Big]-\P_{n,m}\Big[\sum_{s=k+1}^{2k-2}|\C_{s,s}(G)|>n^{(2k-1)(k+1)\alpha}\Big]}{\P_{n,m}(A_k)}\no\\
&&\\
&=&1-O(e^{-n^{(k+1)\alpha}+O(n^{k\alpha})})=1-O(e^{-n^{k\alpha}})\,.\no
\end{eqnarray*}
This finishes proof of Lemma \ref{lem:H-isAll} $\square$
\endproof

\proof{Proof of Lemma \ref{lem:Si-bound}}

Clearly $N_{r,\ell}^{G_t,ij}$ is bounded from above by the number of paths (not necessarily simple paths) of length $r-1$ from $i$ to $j$ that have at least $\ell$
edges of the $G_t$. Number of all such paths is equal to the number of sequences $C=(i=i_0,i_1,\ldots,i_{r-1}=j)$ with $i_{s}\in[n]$ for all $s$, and at least $\ell$ of pairs $(i_{s}i_{s+1})$ in $G_t$. Since $\ell<r-1$ there is a pair $(i_{s}i_{s+1})$ that does not belong to $G_t$.
We take $s$ to be the smallest such number. So any path $C$ breaks into $C=C_1\cup \{(i_{s}i_{s+1})\}\cup C_2$ where $C_1$ is a path starting from $i$ with length $s$ and completely lies inside $G_t$. Number of such paths is at most $\Delta(G)^s$. Similarly $C_2$ is a path with one endpoint equal to $j$ and length $r-2-s$ that contains exactly $\ell-s$ edges of $G_t$. Number of such paths is at most $\Delta(G)^{\ell-s}n^{r-2-\ell}$. Therefore using $G\in \subG_{n,m,k}$,
\begin{equation}
\label{eq:bd-N_rl}
N_{r,\ell}^{G_t,ij}\leq \sum_{s=0}^{\ell}\Delta(G)^\ell n^{r-2-\ell}=O(n^{r-2-\ell+(k+3)\ell\alpha})\,,
\end{equation}
which finishes proof of part (a).

Proof of part (b) is similar. If $s=0$ then clearly the bound $O(n^r)$ is valid since it is the order of all cycles of length $r$. Otherwise, each cycle in $\C_{r,s}$ contains an edge $(ij)\in G$. So the cycle contains a path of length $r$ that contains $(ij)$ and exactly $s-1$ edges of $G\setminus\{(ij)\}$.
Therefore, the number of such cycles is at most $O(\sum_{(ij)\in G}N_{r,s-1}^{G\setminus\{(ij)\},(ij)})$. Note that each cycle is counted at most $s$ times in the bound which is a constant and can be ignored. Using part (a), this number is of order $O(m\Delta(G)^{s-1}n^{r-s-1})=O(\Delta(G)^{s-1}n^{r-s+\alpha})$ which finishes the proof (b).

$\square$
\endproof


\proof{Proof of Lemma \ref{lem:t/m-rplace-theta}} Note that $G_t^\pi$ is a random subgraph of $G$ that has $t$ edges.
Therefore,
\begin{multline*}
\P_{\pi}\le(A_{e_1,\ldots,e_a}^{t,\pi}\cap B_{e_{a+1},\ldots,e_{a+b}}^{t,\pi}\ri)=\frac{{m-a-b\choose t-a}}{{m\choose t}}\\
=\left[\frac{m^{a+b}}{m\cdots(m-a-b+1)}\ri]\le[\frac{(m-t)\cdots(m-t-b+1)}{(m-t)^b}\ri]\le[\frac{t\cdots(t-a+1)}{t^a}\right]f_{a,b}(t)
\end{multline*}
where $f_{a,b}(t)=(\frac{t}{m})^a(\frac{m-t}{m})^b$. This means,
\begin{align}
\P_{\pi}\le(A_{e_1,\ldots,e_a}^{t,\pi}\cap B_{e_{a+1},\ldots,e_{a+b}}^{t,\pi}\ri)(1-\frac{t}{m})^c
&\leq\left(1+\frac{a+b}{m-a-b}\right)^{a+b}f_{a,b+c}(t)\no\\
&\leq \le(1+O(\frac{1}{m})\ri)f_{a,b+c}(t)\,.\label{eq:P_ab<f_ab}
\end{align}
Now using the fact that the function $\theta^a(1-\theta)^b$ has at most one maximum in the interval $(0,1)$ then
\begin{equation}
\label{eq:sumf_ab<int}
\frac{\sum_{t=0}^{m-1}f_{a,b+c}(t)}{m}\leq \int_{\theta=0}^1\theta^a(1-\theta)^{b+c}\,d\theta+O(\frac{1}{m})\,.
\end{equation}
Combining Eqs. \eqref{eq:P_ab<f_ab} and \eqref{eq:sumf_ab<int} proves part (a) of Lemma \ref{lem:t/m-rplace-theta}.

Part (b) is now easy to prove. In particular, given that
\[
\P_{\pi}\le(A_{e_1,\ldots,e_a}^{t,\pi}\cap B_{e_{a+1},\ldots,e_{a+b}}^{t,\pi}\cap
C_{e_{a+b+1}}^{t,\pi}\ri)(1-\frac{t}{m})^c=\frac{{m-a-b-1\choose t-a}}{(m-t){m\choose t}}(1-\frac{t}{m})^c\,,
\]
using a similar bound as above, but with an extra $m$ in the denominator, we have
\begin{eqnarray}
\sum_{t=0}^{m-1}\P_{\pi}\le(A_{e_1,\ldots,e_a}^{t,\pi}\cap
B_{e_{a+1},\ldots,e_{a+b}}^{t,\pi}\cap C_{e_{a+b+1}}^{t,\pi}\ri)(1-\frac{t}{m})^c
&\leq&O(\frac{1}{m})+\frac{\sum_{t=0}^{m-1}f_{a,b+c}(t)}{m}\,,\no
\end{eqnarray}
which finishes proof of part (b) via Eq. \eqref{eq:sumf_ab<int}.

Now, we prove part (c). First we use Bernoulli's inequality $(1-x)^y\geq1-yx$ for $0\leq x<1$, $y\geq 1$
to show that for $m-\sqrt{m}>t>\sqrt{m}$
\begin{align}
\P_{\pi}\le(A_{e_1,\ldots,e_a}^{t,\pi}\cap B_{e_{a+1},\ldots,e_{a+b}}^{t,\pi}\ri)(1-\frac{t}{m})^c
&=(1-\frac{t}{m})^c\frac{{m-a-b\choose t-a}}{{m\choose t}}\no\\
&\geq(1-\frac{a}{t})^a(1-\frac{b}{m-t})^bf_{a,b+c}(t)\no\\
&\geq \le[1-O(\frac{1}{\sqrt{m}})\ri]f_{a,b+c}(t)\,.
\label{P_a>f_a0}
\end{align}
Also, as before,
\begin{equation}
\label{eq:sumf_a0>int}
\frac{\sum_{t=0}^{m-1}f_{a,b+c}(t)}{m}\geq \int_{\theta=0}^1\theta^a(1-\theta)^{b+c}\,d\theta-O(\frac{1}{m})\,.
\end{equation}
Hence,
\begin{eqnarray}
\sum_{t=0}^{m-1}\P_{\pi}\le(A_{e_1,\ldots,e_a}^{t,\pi}\cap B_{e_{a+1},\ldots,e_{a+b}}^{t,\pi}\ri)(1-\frac{t}{m})^c
&\geq&\sum_{\sqrt{m}<t<m-\sqrt{m}}\P_{\pi}\le(A_{e_1,\ldots,e_a}^{t,\pi}\cap B_{e_{a+1},\ldots,e_{a+b}}^{t,\pi}\ri)(1-\frac{t}{m})^c\no\\
&\geq&\le(1-O(\frac{1}{\sqrt{m}})\ri)\sum_{\sqrt{m}<t<m-\sqrt{m}}f_{a,b+c}(t)\no\\
&\geq&\le(1-O(\frac{1}{\sqrt{m}})\ri)\sum_{t=0}^{m-1}f_{a,b+c}(t)-O(\sqrt{m})\no\\
&\geq&\le(m-O(\sqrt{m})\ri)\int_{\theta=0}^1\theta^a(1-\theta)^{b+c}\,d\theta-O(\sqrt{m})\no\\
&=&m\int_{\theta=0}^1\theta^a(1-\theta)^{b+c}\,d\theta-O(\sqrt{m}) \,,\no
\end{eqnarray}
which finishes proof of Lemma \ref{lem:t/m-rplace-theta} $\square$

\endproof

%
%
\proof{Proof of Lemma \ref{lem:p(ij|G_t)}}
The main idea is that each entry of the matrix $\bM_t+\frac{m-t}{{n\choose2}-t}\bM_t^{(c)}$ corresponds to sum of all products of entries of the matrix $\bM_t+\frac{m-t}{{n\choose2}-t}\bM_t^{(c)}$ that correspond to paths of length $r$ in $K_n$. Moreover the sum is dominated by those products that correspond to simple paths rather than self intersecting paths. Below, we will show this formally.

By definition, for any non-zero $(ij)$ entry of the matrix $\bP_{G_t}'$ we have:
\begin{eqnarray}
(\bP_{G_t}')_{ij}& 
=& \exp\le(-\sum_{r=3}^{k}\sum_{\ell=0}^{r-2}N_{r,\ell}^{G_t,ij}q_t^{r-1-\ell} -\sum_{r=3}^{k}\sum_{\ell=0}^{r-2}M_{r,\ell}^{G_t,(ij)}q_t^{r-1-\ell}\ri)\no
\end{eqnarray}
where $M_{r,\ell}^{G_t,(ij)}$ is the number of self intersecting cycles of length $r$ in $K_n$ that include $(ij)$ and exactly $\ell$ edges of $G_t$. Similarly to the argument used in Lemma \ref{lem:Si-bound} to prove an upper bound for $N_{r,\ell}^{G_t,ij}$, we can show that
\begin{equation}
\label{eq:bd-M_rl}
M_{r,\ell}^{G_t,ij} = O(n^{r-3-\ell+(k+3)\ell\alpha})\,.
\end{equation}
There is one factor $n$ less in the right hand side of Eq. \eqref{eq:bd-M_rl} compared to the bound we showed in Lemma \ref{lem:Si-bound} for $N_{r,\ell}^{G_t,ij}$ and the reason is, due to self-intersection of the paths, there exist one less degree of freedom.
Therefore,
\begin{eqnarray}
(\bP_{G_t}')_{ij}& = &\exp\le(-\sum_{r=3}^{k}\sum_{\ell=0}^{r-2}N_{r,\ell}^{G_t,(ij)}q_t^{r-1-\ell} - O\le(n^{k(k+3)\alpha-2}\ri)\ri)\,.\no
\end{eqnarray}
For simplicity of the notation let $D_{ij}^{G_t}=\exp\le(-\sum_{r=3}^{k}\sum_{\ell=0}^{r-2}N_{r,\ell}^{G_t,(ij)}q_t^{r-1-\ell}\ri)$.
Hence,
\begin{eqnarray}
p'(ij|G_t)&=&\frac{(\bP_{G_t}')_{ij}}{Z'(G_t)}\no\\
&=&\frac{(\bP_{G_t}')_{ij}}{\sum_{rs\in Q(G_t)}(\bP_{G_t}')_{rs}}\no\\
&&\\
&=&\frac{D_{ij}^{G_t}\exp\le(- O\le(n^{k(k+3)\alpha-2}\ri)\ri)}
{\sum_{rs\in Q(G_t)}D_{rs}^{G_t}\exp\le(-O\le(n^{k(k+3)\alpha-2}\ri)\ri)}\no\\
&&\\
&\geq&\frac{D_{ij}^{G_t}}
{\sum_{rs\in Q(G_t)}D_{rs}^{G_t}}\exp\le(- O\le(n^{k(k+3)\alpha-2}\ri)\ri)\no
\end{eqnarray}
which finishes the proof $\square$
\endproof

%
%
\proof{Proof of Corollary \ref{cor:Q(G)BoundByUs}}

%
%
Recall from \S \ref{sec:analysis} that $F(G)$ is the number of edges in $K_n$ that when added to $G$ a cycle of length at most $k$ is created.  Clearly, $Q(G)=N-m-F(G)$. On the other hand, it is clear that $F(G)\leq \sum_{r=3}^k |\C_{r,r-1}|$. Therefore, using Lemma \ref{lem:Si-bound}(b), for all $G$ in $\subG_{n,m,k}$
\[
F(G)=O(n^{(k-1)(k+3)\alpha+1})\,.
\]
Hence,
\begin{align*}
1-\frac{|Q_k(G)|}{\E_{\unif|Q_k(G)|}}\leq 1-\frac{N-m-O(n^{(k-1)(k+3)\alpha+1})}{N-m}= \frac{O(n^{(k-1)(k+3)\alpha+1})}{N-m}=O(n^{(k-1)(k+3)\alpha-1})\,.
\end{align*}
Similarly,
\begin{align*}
1-\frac{|Q_k(G)|}{\E_{\unif|Q_k(G)|}}\geq 1-\frac{N-m}{N-m-O(n^{(k-1)(k+3)\alpha+1})}= -\frac{O(n^{(k-1)(k+3)\alpha+1})}{O(n^2)}=-O(n^{(k-1)(k+3)\alpha-1})\,.
\end{align*}
Therefore, combining the above two equations and using Lemma \ref{lem:H-isAll}, there is a constant $c_3$ such that
\begin{align}
\P_{\unif}\left\{\left|1-\frac{|Q_k(G)|}{\E_{\unif|Q_k(G)|}}\right| < c_3 n^{(k-1)(k+3)\alpha-1}\right\}=1-O(e^{-n^{k\alpha}})\,.
\label{eq:Q(G)BoundForUniform}
\end{align}
Now, define the event $A$ by
\[
A= \left\{\left|1-\frac{|Q_k(G)|}{\E_{\unif|Q_k(G)|}}\right| > c_3 n^{(k-1)(k+3)\alpha-1}\right\}\,.
\]
From the definition of $\dv$ and Theorem \ref{thm:main} we have
\begin{align*}
|\P_{\alg}(A)-\P_{\unif}(A)| &\leq \dv(\P_\alg,\P_\unif)=O(n^{-1/2+k(k+3)\alpha})\,.
\end{align*}
Therefore, combining this with Eq. \eqref{eq:Q(G)BoundForUniform},
\begin{align*}
\P_{\alg}(A)&\leq \P_{\unif}(A)+O(n^{-1/2+k(k+3)\alpha})
=O(e^{-n^{k\alpha}})+O(n^{-1/2+k(k+3)\alpha})=O(n^{-1/2+k(k+3)\alpha})\,
\end{align*}
which finishes the proof $\square$
\endproof

\section{Mathematical Notations}\label{app:notations}
{\footnotesize
\begin{longtable}{ll}
		{\bf Notation} & {\bf Description} \\
\hline\\
        $[n]$:& When $n$ is a positive integer it denotes the set $\{1,2,\ldots,n\}$. \\
        $K_n$:& Complete graph with vertex set $[n]$.\\
        $O$:& For sequences $\{a_n\}_{n\ge 1}, \{b_n\}_{n\ge 1}$ big $O$ notation $a_n=O(b_n)$ means $\lim\sup_{n\to\infty}a_n/b_n < \infty$.\\
        $o$:&For sequences $\{a_n\}_{n\ge 1}, \{b_n\}_{n\ge 1}$ little $O$ notation $a_n=o(b_n)$ means $\lim\sup_{n\to\infty}a_n/b_n =0$.\\
        $(ij)$:&An edge that connects node $i$ to node $j$ ($i,j\in[n]$) (in a graph $G$ with vertices $[n]$).\\
        $n$:& Number of vertices of graphs considered in the paper.\\
        $m$:& Number of edges of most graphs in the paper.\\
        $N$:& Defined to be ${n\choose2}$.\\
        $\nedge(G)$:& Number edges of a graph $G$.\\
        $\gens_{n,m}$:&Set of all simple graphs with $m$ edges and vertices $[n]$.\\
        $\gens_{n,p}$:& Random graph model of simple graphs on $[n]$ where each edge is present (independently) with probability $p$.\\
        $\P_{n,m}$:&Uniform probability distribution over $\gens_{n,m}$.\\
        $\P_{n,p}$:&Probability distribution obtained by random graph model $\gens_{n,p}$.\\
        $\gens_{n,m,k}$:& The subset of graphs in $\gens_{n,m}$  with girth greater than $k$.\\
        $\subG_{n,m,k}$:&The set of graphs $G$ in $\gens_{n,m,k}$ with maximum degree of order $O(n^{(k+3)\alpha})$\\
        $\gens_{n,m,k}(\tau)$:&Subset of graphs $G$ in $\gens_{n,m,k}$ where $\P_\alg(G)<(1-\tau)\P_{\unif}(G)$.\\
        $\P_{\alg}$:&Output distribution of $\algname$ which is a distribution on $\gens_{n,m,k}$.\\
        $\P_{\unif}$:&Uniform distribution on $\gens_{n,m,k}$.\\
        $\dv(\P,\mb{Q})$:&Total variation distance between measures on $X$ and is equal to $\sup\,\{|\P(A)-\mb{Q}(A)|\,:\,A\subset X\}$.\\
        $G_t$:&Partially constructed graph in $\algname$ after $t$ steps.\\
        $q_t$:& Equals to $(m-t)/(N-t)$.\\
        $\theta$:&Equals to $t/m$.\\
        $\pi$:& A permutation of the edges of $G$ where $G\in\gens_{n,m}$.\\
        $G_t^\pi$:& The graph having $[n]$ as vertex set and $\{\pi(1),\dots,\pi(t)\}$ as edge set.\\
        $\E_\pi$:& Expectation with respect to a uniformly random permutation $\pi$.\\
        $\P_\pi$:& Probability with respect to a uniformly random permutation $\pi$.\\
        $\gamma$:&Notation used for cycles.\\
        $Q(G_t)$:&The set of edges $(ij)$ that do not belong to $G_t$ and $G_t\cup(ij)\in\gens_{n,t+1,k}$. \\
        $p(ij|G_t)$:& For each $(ij)\in Q(G_t)$, it is the probability of selecting $(ij)$ in step $t$ of $\algname$.\\
        $E_k(G_t,ij)$:&Equals to $\sum_{r=3}^k\sum_{\ell=0}^{r-2}N_{r,\ell}^{G_t,ij}q_t^{r-1-\ell}$.\\
        $\mrm{T}$:&Execution tree of a sequential graph generation algorithm like $\algname$ (see \S \ref{sec:idea} for details).\\
        $\pi_t$:& For a partially constructed graph $G_t$, it is an ordering (permutation) of its edges.\\
        $n_k(G_t,\pi_{t})$:&Number of cycles of length at most $k$ in a random extension of of a pair $(G_t,\pi_t)$ in $\mrm{T}$.\\
        $N_{r,\ell}^{G,ij}$:&Number of simple cycles in $K_n$ that have length $r$, include $(ij)$, and include exactly $\ell$ edges of $G$.
        \\ 
        $Z(G)$:& Normalization constant in definition of $p(ij|G_t)$ in Eq. \eqref{eq:P(G_t)}.\\
        $F(G_t^\pi)$:&The set of edges $(ij)$ where $G_t^\pi\cup(ij)$ has a cycle of length at most $k$.\\
        $Z_0(G)$:& Is equal to $N-t-F(G_t^\pi)$.\\
        $S_1(G)$:&Equals to $-\sum_{t=0}^{m-1}\E_\pi E_k(G_t^\pi,\pi(t+1))$.\\
        $S_2(G)$:&Equals to $\frac{1}{N}\sum_{t=0}^{m-1} \E_{\pi} F(G^{\pi}_t)$.\\
        $S_3(G)$:&Equals to $-\sum_{t=0}^{m-1}\E_{\pi}\log \frac{Z(G_t^{\pi})}{Z_0(G_t^{\pi})}$.\\
        $\C_{r}$:& Set of all simple cycles of length $r$ in $K_n$.\\
        $\C_{r,\ell}(G)$:& Cycles in $\C_{r}$ that include exactly $\ell$ edges of $G$.\\
        $\gamma_{r,s}$:& An element of $\C_{r,\ell}(G)$.\\
        $s_i(C_{r,s})$:& For each $i=1,2,3$ denotes contribution of cycle $C_{r,s}$ in $S_i(G)$.\\
        $A_k$:& The event that a random graph has girth greater than $k$.\\
        $\deg_v(H)$:& Induced degree of a note $v$ in a subgraph $H$ of a larger graph containing $v$.\\
        $\Delta(G)$:& Maximum degree of graph $G$.\\
        $A_{e_1,\ldots,e_s}^{t,\pi}$:& The event $\{\forall i\in [s]~:~e_i\in G_t^\pi\}$ when $e_1,\ldots,e_s$ are edges of $G$.\\
        $B_{e_1,\ldots,e_s}^{t,\pi}$:& The event $\{\forall i\in[s]~:~e_i\notin G_t^\pi\}$ when $e_1,\ldots,e_s$ are edges of $G$.\\
        $C_{e}^{t,\pi}$:& The event $\{\pi(t+1)=e\}$ for edge $e$ in $G$.\\
        $\bM_{t}$:& Adjacency matrix of $G_t$.\\
        $\bM_{t}^{(c)}$:& Adjacency matrix of complement of $G_t$.\\
        $\bQ_{t}$:& Adjacency matrix of all edges in $Q(G_t)$.\\
        $\mathbf{A}=\mathbf{B}\odot \mathbf{C}$:& For $n\times n$ matrices $\mathbf{A}, \mathbf{B}, \mathbf{C}$ it means that for all $i,j\in [n]$:~~$a_{ij}=b_{ij}c_{ij}$.\\
        $\mathbf{A}=\widehat{\exp}(\mathbf{B})$:& For $n\times n$ matrices $\mathbf{A}, \mathbf{B}$ it means that for all $i,j\in [n]$:~~$a_{ij}=e^{b_{ij}}$.\\
        $\mathbf{A}=\widehat{{\rm sign}}(\mathbf{B})$:& For $n\times n$ matrices $\mathbf{A}, \mathbf{B}$ it means that for all $i,j\in [n]$:~~$a_{ij}={\rm sign}(b_{ij})$.\\
        $\mathbf{J}_n$:& It is the $n$ by $n$ matrix of all ones.\\
		  \hline
\caption{Mathematical notations.}
\label{tab:notations}
\end{longtable}
}
\end{APPENDICES}

\ACKNOWLEDGMENT{
The authors gratefully acknowledge the National Science Foundation (awards CMMI: 1554140 and CCF: 1216698) and Office of Naval Research (N00014-16-1-2893) for financial support.

This paper has also benefitted from valuable feedback from Balaji Prabhakar, Joel Spencer, Daniel Spielman, Stefanos Zenios, and anonymous referees.
}
\vspace{5mm}

\bibliographystyle{ormsv080} 
\bibliography{references}

\begin{thebibliography}{46}
\expandafter\ifx\csname natexlab\endcsname\relax\def\natexlab#1{#1}\fi
\expandafter\ifx\csname url\endcsname\relax
  \def\url#1{{\tt #1}}\fi
\expandafter\ifx\csname urlprefix\endcsname\relax\def\urlprefix{URL }\fi
\expandafter\ifx\csname urlstyle\endcsname\relax
  \expandafter\ifx\csname doi\endcsname\relax
  \def\doi#1{doi:\discretionary{}{}{}#1}\fi \else
  \expandafter\ifx\csname doi\endcsname\relax
  \def\doi{doi:\discretionary{}{}{}\begingroup \urlstyle{rm}\Url}\fi \fi

\bibitem[{Alon and Spencer(1992)}]{AlS92}
Alon, N., J.~Spencer. 1992.
\newblock {\it The Probabilistic Method\/}.
\newblock Wiley, New York.

\bibitem[{Amraoui et~al.(2007)Amraoui, Montanari, and Urbanke}]{AMU06}
Amraoui, A., A.~Montanari, R.~Urbanke. 2007.
\newblock How to find good finite-length codes: From art towards science.
\newblock {\it Eur. Trans. Telecomm.\/} {\bf 18} 491--508.

\bibitem[{Bayati et~al.(2009{\natexlab{a}})Bayati, Keshavan, Montanari, Oh, and
  Saberi}]{ITWpaper}
Bayati, M., R.~Keshavan, A.~Montanari, S.~Oh, A.~Saberi. 2009{\natexlab{a}}.
\newblock Generating random tanner graphs with large girth.
\newblock {\it IEEE Information Theory Workshop\/}. Taormina, Italy.
\newblock Code available here:
  \url{http://web.engr.illinois.edu/~swoh/software/girth/index.html}.

\bibitem[{Bayati et~al.(2010)Bayati, Kim, and Saberi}]{BKS07}
Bayati, Mohsen, Jeong~Han Kim, Amin Saberi. 2010.
\newblock A sequential algorithm for generating random graphs.
\newblock {\it Algorithmica\/} {\bf 58}(4) 860--910.

\bibitem[{Bayati et~al.(2009{\natexlab{b}})Bayati, Montanari, and
  Saberi}]{SODA_Version}
Bayati, Mohsen, Andrea Montanari, Amin Saberi. 2009{\natexlab{b}}.
\newblock Generating random graphs with large girth.
\newblock {\it Proceedings of the Twentieth Annual ACM-SIAM Symposium on
  Discrete Algorithms\/}. SODA '09, 566--575.
\newblock \urlprefix\url{http://dl.acm.org/citation.cfm?id=1496770.1496833}.

\bibitem[{Bender and Canfield(1978)}]{BenderC78}
Bender, Edward~A., E.~Rodney Canfield. 1978.
\newblock The asymptotic number of labeled graphs with given degree sequences.
\newblock {\it J. Comb. Theory, Ser. A\/} {\bf 24}(3) 296--307.

\bibitem[{Blanchet(2009)}]{blanchet}
Blanchet, J. 2009.
\newblock Efficient importance sampling for binary contingency tables.
\newblock {\it Ann. Appl. Probab.\/} {\bf 19} 949--982.

\bibitem[{Blitzstein and Diaconis(2010)}]{JoePersi}
Blitzstein, J., P.~Diaconis. 2010.
\newblock A sequential importance sampling algorithm for generating random
  graphs with prescribed degrees.
\newblock {\it Internet Math.\/} {\bf 6} 489--522.

\bibitem[{Bohman and Keevash(2010)}]{BohmanKeevash}
Bohman, T., P.~Keevash. 2010.
\newblock The early evolution of the $h$-free process.
\newblock {\it Inventiones mathematicae\/} {\bf 181}(2) 291--336.

\bibitem[{{Bohman} and {Keevash}(2013)}]{BohmanKeevash2013}
{Bohman}, T., P.~{Keevash}. 2013.
\newblock {Dynamic concentration of the triangle-free process}
  \urlprefix\url{https://arxiv.org/abs/1302.5963}.

\bibitem[{Bollob\'{a}s and Riordan(2000)}]{BoR00}
Bollob\'{a}s, B., O.~Riordan. 2000.
\newblock Constrained graph processes.
\newblock {\it Electronic Journal of Combinatorics\/} {\bf 7}.

\bibitem[{Bollob{\'a}s(1980)}]{bollobas80}
Bollob{\'a}s, B{\'e}la. 1980.
\newblock A probabilistic proof of an asymptotic formula for the number of
  labelled regular graphs.
\newblock {\it European Journal of Combinatorics\/} {\bf 1}(4) 311--316.

\bibitem[{Bu and Towsley(2002)}]{bu}
Bu, T., D.~Towsley. 2002.
\newblock On distinguishing between internet power law topology generators.
\newblock {\it INFOCOM\/}. IEEE.

\bibitem[{Chandrasekhar(2015)}]{Chandrasekhar}
Chandrasekhar, A. 2015.
\newblock Econometrics of network formation.
\newblock Oxford handbook on the economics of networks. (edited by yann
  bramoulle, andrea galeotti and brian rogers).

\bibitem[{Chen et~al.(2005)Chen, Diaconis, Holmes, and
  Liu}]{ChenDiaconisHolmsLiu}
Chen, Y., P.~Diaconis, S.~Holmes, J.~S. Liu. 2005.
\newblock Sequential monte carlo methods for statistical analysis of tables.
\newblock {\it Journal of the American Statistical Association\/} {\bf 100}
  109--120.

\bibitem[{Chung et~al.(2001)Chung, Forney, Richardson, and Urbanke}]{Chung}
Chung, S.~Y., G.~D. Forney, T.~J. Richardson, R.~Urbanke. 2001.
\newblock On the design of low-density parity-check codes within $0.0045$ db of
  the shannon limit.
\newblock {\it IEEE Comm.~Lett\/} {\bf 5} 58--60.

\bibitem[{Di et~al.(2002)Di, Proietti, Teletar, Richardson, and
  Urbanke}]{Stopping}
Di, C., D.~Proietti, I.~E. Teletar, T.~J. Richardson, R.~Urbanke. 2002.
\newblock Finite-length analysis of low-density parity-check codes on the
  binary erasure channel.
\newblock {\it IEEE Trans. Inform. Theory\/} {\bf 46}.

\bibitem[{Efron(1979)}]{efron79}
Efron, B. 1979.
\newblock Bootstrap methods: another look at the jackknife.
\newblock {\it Ann. Statistics\/} {\bf 7} 1--26.

\bibitem[{Erd\H{o}s et~al.(1995)Erd\H{o}s, Suen, and Winkler}]{ESW95}
Erd\H{o}s, P., S.~Suen, P.~Winkler. 1995.
\newblock On the size of a random maximal graph.
\newblock {\it Random Structure and Algorithms\/} {\bf 6} 309--318.

\bibitem[{Faloutsos et~al.(1999)Faloutsos, Faloutsos, and
  Faloutsos}]{faloutsos}
Faloutsos, M., P.~Faloutsos, Ch. Faloutsos. 1999.
\newblock On power-law relationships of the internet topology.
\newblock ACM, New York, NY, USA, 251--262.

\bibitem[{Ioannides(2006)}]{Ioannides2006}
Ioannides, Y. 2006.
\newblock Random graphs and social networks: An economics perspective.
\newblock Preprint.

\bibitem[{Jackson and Watts(2002)}]{JacksonWatts2002}
Jackson, M., D.~Watts. 2002.
\newblock The evolution of social and economic networks.
\newblock {\it Journal of Economic Theory\/} {\bf 106} 265--295.

\bibitem[{Janson et~al.(2000)Janson, {\L}uczak, and Rucinski}]{JLR00}
Janson, {\L}uczak, Rucinski. 2000.
\newblock {\it Random Graphs\/}.
\newblock Wiley-Interscience.

\bibitem[{Janson(1990)}]{Janson90}
Janson, S. 1990.
\newblock Poisson approximation for large deviations.
\newblock {\it Random Structures and Algorithms\/} {\bf 1} 221–229.

\bibitem[{Kim and Vu(2007)}]{Kim-Vu}
Kim, J.~H., V.~H. Vu. 2007.
\newblock Generating random regular graphs.
\newblock {\it Combinatorica\/} {\bf 26} 683--708.

\bibitem[{Kleinberg(2000)}]{Kleinberg2000}
Kleinberg, J. 2000.
\newblock Navigation in a small world.
\newblock {\it Nature\/} {\bf 406} 845.

\bibitem[{Koetter and Vontobel(2003)}]{Pseudo}
Koetter, R., P.~Vontobel. 2003.
\newblock Graph covers and iterative decoding of finite-lenght codes.
\newblock {\it Proc. Int. Conf. on Turbo codes and Rel. Topics\/}. Brest,
  France.

\bibitem[{Luby et~al.(1997)Luby, Mitzenmacher, Shokrollahi, Spielman, and
  Stemann}]{LubyEtAl}
Luby, M., M.~Mitzenmacher, A.~Shokrollahi, D.~A. Spielman, V.~Stemann. 1997.
\newblock Practical loss-resilient codes.
\newblock {\it ACM Symposium on Theory of Computing (STOC)\/}.

\bibitem[{Medina et~al.(2000)Medina, Matta, and Byers}]{medina}
Medina, A., I.~Matta, J.~Byers. 2000.
\newblock On the origin of power laws in internet topologies.
\newblock {\it ACM Computer Communication Review\/} {\bf 30} 18--28.

\bibitem[{Milo et~al.(2002)Milo, ShenOrr, Itzkovitz, Kashtan, Chklovskii, and
  Alon}]{MiloShenOrrItzkovitzKashtanChklovskiiAlon}
Milo, R., S.~ShenOrr, S.~Itzkovitz, N.~Kashtan, D.~Chklovskii, U.~Alon. 2002.
\newblock Network motifs: Simple building blocks of complex networks.
\newblock {\it Science\/} {\bf 298} 824--827.

\bibitem[{Newman(2003)}]{Newman2003}
Newman, M. 2003.
\newblock The structure and function of complex networks.
\newblock {\it SIAM Review\/} {\bf 45} 167--256.

\bibitem[{Osthus and Taraz(2001)}]{OsT99}
Osthus, D., A.~Taraz. 2001.
\newblock Random maximal h-free graphs.
\newblock {\it Random Struct. Algorithms\/} {\bf 18}(1) 61--82.

\bibitem[{Papadimitriou(2001)}]{Papadimitriou2003}
Papadimitriou, C. 2001.
\newblock Algorithms, games, and the internet  749--753.

\bibitem[{Pontiveros et~al.(2013)Pontiveros, Griffiths, and
  Morris}]{Morris2013}
Pontiveros, G.~F., S.~Griffiths, R.~Morris. 2013.
\newblock The triangle-free process and $r(3,k)$.
\newblock \urlprefix\url{http://arxiv.org/abs/1302.6279}.
\newblock Eprint.

\bibitem[{Richardson(2003)}]{Trapping}
Richardson, T. 2003.
\newblock Error-floors of ldpc codes.
\newblock {\it Proceedings of the 41st Annual Conference on Communication,
  Control and Computing\/}. 1426--1435.

\bibitem[{Richardson and Urbanke(2008)}]{MCT}
Richardson, T., R.~Urbanke. 2008.
\newblock {\it Modern Coding Theory\/}.
\newblock Cambridge University Press, Cambridge.

\bibitem[{Rucinski and Wormald(1992)}]{RuW92}
Rucinski, A., N.~Wormald. 1992.
\newblock Random graph processes with degree restrictions.
\newblock {\it Combinatorics Prob. Comput.\/} {\bf 1}.

\bibitem[{Sinclair(1993)}]{SinclairBook}
Sinclair, A. 1993.
\newblock {\it Algorithms for random generation and counting: a Markov chain
  approach\/}.
\newblock Birkhauser.

\bibitem[{Spencer(1995)}]{Spe95}
Spencer, J. 1995.
\newblock Maximal triangle-free graphs and ramsey $r(3,t)$.
\newblock Manuscript.

\bibitem[{Steger and Wormald(1999)}]{StW99}
Steger, A., N.~C. Wormald. 1999.
\newblock Generating random regular graphs quickly.
\newblock {\it Combinatorics Prob. and Comput\/} {\bf 8} 377--396.

\bibitem[{Tangmunarunkit et~al.(2002)Tangmunarunkit, Govindan, Jamin, Shenker,
  and Willinger}]{inet}
Tangmunarunkit, H., R.~Govindan, S.~Jamin, S.~Shenker, W.~Willinger. 2002.
\newblock Network topology generators: Degree-based vs. structural.
\newblock {\it Proceedings of the 2002 Conference on Applications,
  Technologies, Architectures, and Protocols for Computer Communications\/}.
  SIGCOMM '02, ACM, New York, NY, USA, 147--159.

\bibitem[{Valente et~al.(2009)Valente, Fujimoto, Chou, and
  Spruijt-Metz}]{Obesity}
Valente, T., K.~Fujimoto, C.~Chou, D.~Spruijt-Metz. 2009.
\newblock Adolescent affiliations and adiposity: A social network analysis of
  friendships and obesity.
\newblock {\it J Adolesc Health\/} {\bf 45} 202–204.
\newblock \doi{10.1016/j.jadohealth.2009.01.007}.

\bibitem[{Vu(2002)}]{Vu-extension}
Vu, Van~H. 2002.
\newblock Concentration of non-lipschitz functions and applications.
\newblock {\it Random Struct. Algorithms\/} {\bf 20}(3) 262--316.

\bibitem[{Warnke(2014)}]{Warnke}
Warnke, L. 2014.
\newblock The $c_\ell$-free process.
\newblock {\it Random Struct. Algorithms\/} {\bf 44}(4) 490--526.

\bibitem[{Wolfovitz(2011)}]{Wolfovitz11}
Wolfovitz, G. 2011.
\newblock Triangle-free subgraphs in the triangle-free process.
\newblock {\it Random Struct. Algorithms\/} {\bf 39}(4) 539--543.

\bibitem[{Wormald(1999)}]{Wormald1999}
Wormald, N.~C. 1999.
\newblock Models of random regular graphs.
\newblock {\it London Mathematical Society Lecture Note Series\/}  239--298.

\end{thebibliography}

\end{document}